\documentclass[lettersize,journal]{IEEEtran}

\usepackage{amsmath,amsfonts,amssymb}
\usepackage{graphicx}
\usepackage{textcomp,cite,stfloats,array,tabularx,booktabs,multirow}
\usepackage[table]{xcolor}
\usepackage[hidelinks]{hyperref}

\newcolumntype{C}{>{\centering\arraybackslash}X}
\newcolumntype{R}{>{\raggedleft\arraybackslash}X}
\newcolumntype{L}{>{\raggedright\arraybackslash}X}

\newcommand{\ER}{$\mathrm{ER}_{\le 20^\circ}$}
\newcommand{\Fone}{$\mathrm{F}_{\le 20^\circ}$}
\newcommand{\LE}{$\mathrm{LE}_\mathrm{CD}$}
\newcommand{\LR}{$\mathrm{LR}_\mathrm{CD}$}
\newcommand{\ESELD}{$\mathcal{E}_{\mathrm{SELD}}$}

\begin{document}

\title{MAGENTA: Magnitude and Geometry-Enhanced Training Approach for Long-Tailed Sound Event Localization and Detection}

\author{Jun-Wei Yeow, \IEEEmembership{Student Member, IEEE}, Ee-Leng Tan, \IEEEmembership{Senior Member, IEEE}, Santi Peksi, \IEEEmembership{Member, IEEE}, Woon-Seng Gan, \IEEEmembership{Senior Member, IEEE}
        % <-this % stops a space
\thanks{J.-W. Yeow, E.-L. Tan, S. Peksi, and W.-S. Gan are with the Smart
Nation TRANS Lab, School of Electrical and Electronic Engineering, Nanyang Technological University, 639798, Singapore (email: junwei004@e.ntu.edu.sg, \{etanel, speksi, ewsgan\}@ntu.edu.sg)}
\thanks{This work was supported by the Ministry of Education, Singapore, through Academic Research Fund Tier 2 under Grant MOE-T2EP20224-0010 and Grant MoE Tier-1: RG13/25.}
\thanks{Accepted for publication in the IEEE Transactions on Audio, Speech, and Language Processing.}
\thanks{\textcopyright~2026 IEEE. Personal use of this material is permitted. Permission from IEEE must be obtained for all other uses, in any current or future media, including reprinting/republishing this material for advertising or promotional purposes, creating new collective works, for resale or redistribution to servers or lists, or reuse of any copyrighted component of this work in other works.}}

\maketitle

\begin{abstract}
Deep learning-based Sound Event Localization and Detection (SELD) systems suffer severe performance degradation in real-world, long-tailed acoustic environments. Standard continuous regression objectives heavily bias learning toward frequent classes, causing rare events to be systematically under-recognized, an optimization bottleneck we term \textit{detection timidity}. To overcome this, we propose MAGENTA (Magnitude And Geometry-ENhanced Training Approach), an architecture-agnostic loss framework that geometrically decomposes the regression error into orthogonal radial (activity) and angular (localization) components. Unlike standard methods that rely on static frequency weights, MAGENTA incorporates an intrinsic, difficulty-driven annealing mechanism. By decoupling the objective to independently modulate active detection and inactive suppression, the system can adaptively boost recall for difficult tail classes while modulating inactive penalties to prevent spurious rare-event detections. Evaluations on the STARSS23 dataset demonstrate that MAGENTA yields a 20.5\% relative reduction in the aggregated SELD error, effectively recovering tail class performance without compromising head class precision. Code is available at: \url{https://github.com/itsjunwei/MAGENTA}
\end{abstract}


\section{Introduction}
\label{sec:intro}

\IEEEPARstart{S}{ound} Event Localization and Detection (SELD) is a fundamental task in machine hearing~\cite{virtanen2018computational, mesaros2025decade}, enabling systems to jointly identify acoustic events and estimate their direction-of-arrival (DOA)~\cite{adavanne2018sound}. While recent advancements have expanded SELD to include distance estimation for full 3-D spatial localization~\cite{krause2024sound, yeow2025improving}, this work strictly addresses the angular DOA estimation paradigm. This capability drives situational awareness in real-world applications such as autonomous navigation and environmental monitoring~\cite{yeow2024real, wang2023loss, yeow2025environmental}. Modern deep learning architectures~\cite{hirvonen2015classification} have largely replaced traditional signal processing pipelines to handle these complex, reverberant acoustic scenes. To further improve end-to-end efficiency, state-of-the-art deep learning SELD architectures have converged on the Activity-Coupled Cartesian DOA (ACCDOA) representation~\cite{shimada2021accdoa, shimada2022multi}. Unlike early multi-task formulations~\cite{adavanne2018sound, cao2019polyphonic} that required balancing separate classification and regression heads~\cite{phan2020multitask}, ACCDOA elegantly unifies detection and localization into a single continuous regression vector: the direction encodes the spatial origin, while the magnitude encodes activity probability\footnote{An earlier, preliminary version of this work appears as v1 of this arXiv submission. The present manuscript substantially expands that version through refined mathematical formulations, dynamic annealing mechanisms, and comprehensive evaluations across multiple state-of-the-art architectures.}.

\begin{figure}[t]
    \centering
    \includegraphics[width=0.99\linewidth]{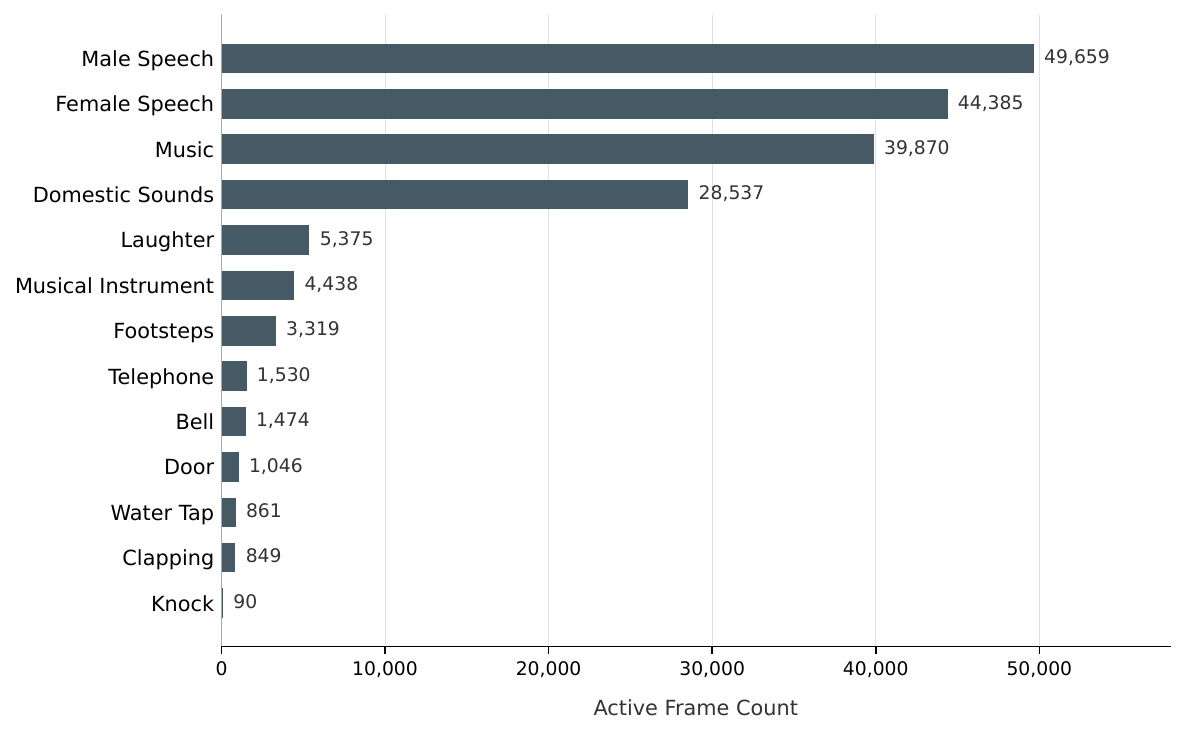}
    \caption{Distribution of active sound event frames in the real-world STARSS23~\cite{shimada2023starss23} training set. The environment is characterized by extreme class imbalance, where frequent ``head'' classes (e.g., \textit{Male Speech}) possess over 550$\times$ more training data than severe ``tail'' classes (e.g., \textit{Knock}).}
    \label{fig:frame_dist}
\end{figure}

However, this efficient coupling creates a unique challenge when applied to real-world datasets characterized by long-tailed distributions~\cite{cao2019LDAM, yang2021delving, zhang2022label}. As illustrated in Fig.~\ref{fig:frame_dist}, real-world acoustic environments, such as those in the Sony-TAu Realistic Spatial Soundscapes 2023 (STARSS23) dataset~\cite{shimada2023starss23}, exhibit severe class imbalance. These distributions are heavily skewed by frequent ``head" classes (e.g., \textit{Male Speech}), whereas critical ``tail" events (e.g., \textit{Knock}) remain exceedingly rare. When optimized with standard regression objectives such as Mean Squared Error (MSE) under extreme imbalance, ACCDOA-based models exhibit a phenomenon we term \textit{\textbf{detection timidity}}.

We formally define this as the \textbf{systematic collapse of prediction magnitudes for minority classes toward zero}, an avoidance behavior driven by the network minimizing the overwhelming volume of inactive-frame penalties. As MSE treats the continuous regression vector as a generic point in Euclidean space, the overwhelming number of silent, zero-norm frames for rare classes dominates the gradient descent process~\cite{shi2023re, Tan_EqualizationLosses_2023, ren2022balanced}. To minimize the global error, the model systematically suppresses prediction magnitudes for tail classes, resulting in negligible recall for these rare events, despite accurate head class predictions~\cite{zhang2023deep, cao2019LDAM, zhang2025systematic, tang2020long}.

Addressing this detection timidity is non-trivial. Standard data-level long-tailed learning (LTL) techniques, such as re-sampling to artificially balance the training distribution~\cite{shi2023re, zhang2022label, chawla2002smote}, fail because real-world acoustic scenes are inherently polyphonic~\cite{politis2022starss22, shimada2023starss23}; attempting to over-sample a rare event inevitably co-samples concurrent, frequent classes~\cite{wu2020distribution}. Similarly, standard algorithmic loss re-weighting is structurally incompatible. Designed primarily for categorical cross-entropy probabilities~\cite{lin2017focal, cui2019class, imoto2022impact}, blindly applying static penalty weights to a coupled spatial vector arbitrarily inflates spatial localization gradients simply to correct an activity confidence deficit. While long-tailed regression is an emerging field~\cite{ren2022balanced, yang2021delving, pu2025leveraging}, existing methods assume the target is a continuous semantic variable (e.g., age) where neighboring targets share feature representations. The imbalance in ACCDOA, however, stems from the discrete frequency of class occurrences mathematically coupled within a continuous, geometric space. 

Furthermore, relying on static weights assumes that statistical rarity strictly equates to learning difficulty~\cite{tian2022striking, park2021influence}. In complex acoustic environments, a rare event may be spectrally distinct and easy to learn, while a frequent event may be highly diffuse~\cite{imoto2022impact}. While statistical re-weighting can artificially improve recall, consistent heavy penalties can force the network to over-predict, thereby severely degrading precision via an influx of false positives. Therefore, a robust SELD solution must accomplish two objectives: it must geometrically decouple the error to isolate active detection from spatial localization, and it must dynamically adapt its learning curriculum to the model's instantaneous performance.

In this work, we propose MAGENTA (Magnitude And Geometry-ENhanced Training Approach), a dynamic loss framework designed specifically for long-tailed ACCDOA-based SELD. MAGENTA geometrically decomposes the regression error into orthogonal radial (activity) and angular (localization) components, isolating active detection from spatial localization. This permits targeted, difficulty-aware priors to independently modulate active detection and inactive suppression. Crucially, by dynamically adapting the penalty modulation based on the model's real-time geometric competence, MAGENTA intrinsically phases out rarity weights once a class is learned, actively resolving the precision-recall trade-off~\cite{tian2022striking} inherent in static LTL. 

To the best of our knowledge, this work is the first to directly address long-tailed class imbalance within the continuous SELD regression space without resorting to synthetic data augmentation. Our primary contributions are as follows:
\begin{enumerate}
    \item We formulate a novel regression objective that geometrically disentangles magnitude errors from angular errors, enabling targeted modulation that mitigates ``detection timidity" caused by zero-norm gradient dominance.
    
    \item We introduce an online, self-pacing modulation mechanism that decouples statistical rarity from acoustic learning difficulty. By dynamically scaling active and inactive losses independently, the framework learns rare events and adapts to difficult classes, successfully resolving the precision-recall trade-off.
    
    \item We evaluate MAGENTA on the highly imbalanced, real-world STARSS23 dataset, demonstrating that it significantly outperforms standard regression baselines and classification-based static re-weighting strategies. 
    
    \item We demonstrate that MAGENTA is an architecture-agnostic framework that easily integrates into state-of-the-art SELD models, yielding consistent performance gains with negligible computational overhead.
\end{enumerate}

\section{Related Work}
\label{sec:related_work}

\subsection{Unified Output Representations for SELD}

As a foundational component of machine hearing, SELD aims to continuously map the ``what" and ``where" of acoustic environments~\cite{yeow2025environmental}. Accelerated by initiatives such as the Detection and Classification of Acoustic Scenes and Events (DCASE) Challenges~\cite{mesaros2025decade}, the community has benefited from publicly available baselines~\cite{krause2024sound, adavanne2018sound}, datasets~\cite{adavanne2019_tnsse2019, politis2020_tnsse2020, politis2021dataset_tnsse2021, politis2022starss22, shimada2023starss23}, and evaluation metrics~\cite{politis2020overview}. Modern investigations have rapidly expanded the operational scope of SELD beyond stationary microphone arrays to encompass highly complex, real-world considerations, including wearable~\cite{nagatomo2022wearable, yasuda20246dof}, stereo~\cite{shimada2025stereo, yeow2025towards}, and emerging multimodal paradigms incorporating visual~\cite{shimada2023starss23, berghi2024fusion} or motion cues~\cite{yasuda20246dof}.

Early deep learning formulations treated detection and localization as disjoint optimization problems~\cite{adavanne2018sound, cao2019polyphonic}. These architectures utilized separate task heads for multi-label sound event detection (SED) and DOA estimation. While effective, multi-branch architectures incurred substantial computational overhead and necessitated complex hyperparameter tuning to balance fundamentally competing loss functions~\cite{phan2020multitask}.

The introduction of the ACCDOA representation~\cite{shimada2021accdoa, shimada2022multi} resolved these inefficiencies by embedding the activity probability directly into the vector's magnitude, streamlining SELD into a unified end-to-end framework optimized via a single continuous regression objective. Consequently, ACCDOA has cemented itself as the \textit{de facto} output representation for both traditional and cutting-edge SELD paradigms~\cite{krause2024sound, shimada2025stereo, shimada2023starss23, yasuda20246dof, nagatomo2022wearable}, achieving state-of-the-art performance across various benchmarks~\cite{shul2024cst, yeow2024squeeze, hu2025pseldnets}. Given the ubiquity of this representation, refining the core loss framework that governs the continuous ACCDOA vector space yields compounding benefits for the broader trajectory of SELD research.

\subsection{Data-Level Mitigation in Polyphonic Contexts}

Mitigating long-tailed distributions is a widely recognized bottleneck in deep learning~\cite{zhang2023deep, zhang2025systematic, yang2022survey, yang2021delving}. Data-level techniques, such as over-sampling minority classes or under-sampling majority classes~\cite{zhang2022label, chawla2002smote}, seek to explicitly balance the training distribution. However, as established, the inherent polyphony of real-world acoustic scenes~\cite{politis2022starss22, shimada2023starss23} renders simple re-sampling ineffective. Attempting to over-sample a rare target inevitably leads to the co-sampling of concurrent, overlapping head classes~\cite{wu2020distribution, ridnik2021asymmetric}, thereby failing to alter the relative class imbalance. 

To alleviate this issue, the SELD community has largely relied on synthetic data augmentation utilizing Room Impulse Responses (RIRs)~\cite{roman2024spatial}. While synthetic pipelines can artificially inflate the volume of tail class samples, they introduce a severe ``domain shift" challenge~\cite{yeow2025_enhancing, hu2025pseldnets}. Conventional synthetic generation toolkits often rely on simplified acoustic models~\cite{scheibler2018pyroomacoustics} that fail to replicate the complex reverberation and noise characteristics of real-world environments~\cite{suzic2024exterior, zhang2025sound}. Furthermore, existing synthetic augmentation methods are largely incapable of replicating dynamic deployment scenarios, such as continuous user head rotation~\cite{yasuda20246dof}. These critical limitations emphasize the need for robust algorithmic loss modulation, which offers a highly scalable alternative to data manipulation by addressing the imbalance directly during gradient optimization~\cite{zhang2023deep, cui2019class}.

\subsection{Algorithmic Loss Modulation for Continuous Spaces}

Core algorithmic methods utilize loss re-weighting, where the gradient penalty is re-balanced based on statistical class-wise distributions~\cite{zhang2024unified, zhang2023deep, wu2020distribution}, with techniques such as Inverse Frequency Loss (IFL)~\cite{imoto2021impact} and Class-Balanced loss~\cite{cui2019class} proving successful in categorical classification tasks. However, applying these standard LTL approaches to the continuous ACCDOA vector presents structural incompatibilities, requiring either reverting to the aforementioned, inefficient two-branch architecture~\cite{phan2020multitask}, or applying scalar weights directly to a coupled spatial vector without geometric consideration. The latter distorts the geometric integrity of the representation, as indiscriminately scaling the error vector arbitrarily inflates spatial localization gradients to correct an activity confidence deficit.

Beyond structural incompatibilities, static frequency-based weighting~\cite{imoto2021impact, imoto2022impact, zhang2024unified, cui2019class} operates on the assumption that statistical rarity dictates learning difficulty~\cite{tian2022striking}. Heavily penalizing rare but distinct classes can force the network to over-predict them, inducing severe precision degradation via false positive hallucinations~\cite{zhang2023deep}. To overcome this precision-recall trade-off, modern solutions have pivoted toward dynamic loss landscapes such as Focal Loss~\cite{lin2017focal, li2022generalizedfocal} for classification or Balanced MSE~\cite{ren2022balanced} for density-based regression. Despite these advancements, there remains a critical gap for a fully dynamic, self-pacing objective tailored specifically to navigate the coupled geometric complexities of SELD. The proposed MAGENTA framework is explicitly formulated to bridge this gap by geometrically decoupling these competing spatial and semantic objectives.

\section{Proposed Method}
\label{sec:theory}

\begin{figure*}[t]
    \centering
    \includegraphics[width=0.99\linewidth]{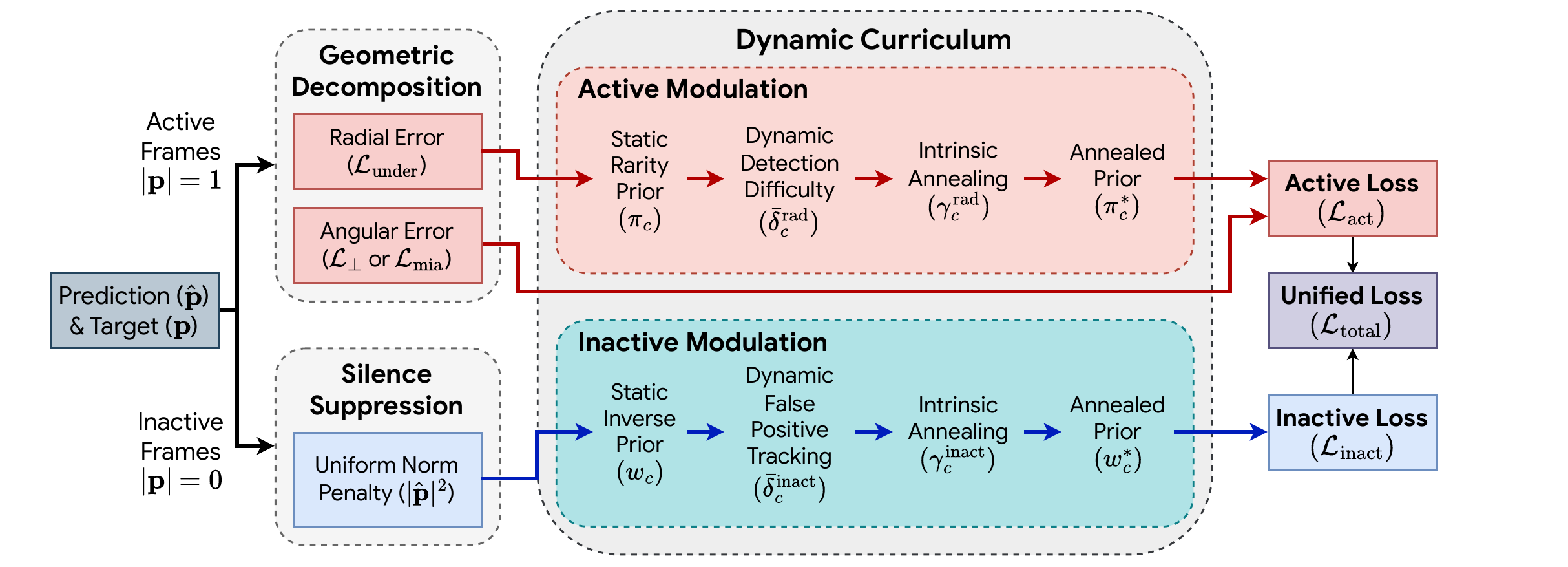}
    \caption{Conceptual flowchart of the proposed MAGENTA framework. The continuous ACCDOA regression space is geometrically decoupled, allowing the objective to process active frames (top, red) and inactive frames (bottom, blue) independently. Across both domains, static statistical priors ($\pi_c, w_c$) are dynamically modulated by the model's instantaneous detection difficulty ($\bar{\delta}_c^{\mathrm{rad}}, \bar{\delta}_c^{\mathrm{inact}}$) and intrinsically annealed ($\gamma_c^{\mathrm{rad}}, \gamma_c^{\mathrm{inact}}$). This targeted curriculum allows the network to recover rare sound events without sacrificing spatial precision or introducing false-positive hallucinations.}
    \label{fig:magenta_flowchart}
\end{figure*}

Before detailing the mathematical foundation, we provide a high-level overview of the MAGENTA framework, as illustrated in Fig.~\ref{fig:magenta_flowchart}. Standard regression objectives inadvertently force models to suppress rare sound events to minimize penalties during silent frames. MAGENTA circumvents this by geometrically decoupling regression errors, allowing the network to actively recover rare sound events without sacrificing spatial precision or introducing spurious false positives. 

For clarity, we formulate our method using the single-track ACCDOA format~\cite{shimada2021accdoa}. Let the ground-truth target for class $c \in \{1, \dots, C\}$ at time frame $t \in \{1, \dots, T\}$ be $\mathbf{p}_{t,c} \in \mathbb{R}^3$. In this representation, the target vector norm determines activity: $\lVert \mathbf{p}_{t,c} \rVert = 1$ if the class is active, and $0$ otherwise. We denote the binary activity label as $y_{t,c} = \mathbb{I} ( \lVert \mathbf{p}_{t,c} \rVert > 0.5)$, where $\mathbb{I}(\cdot)$ is the indicator function that returns 1 if the condition is true and 0 otherwise. The network predicts a vector $\hat{\mathbf{p}}_{t,c}$ with magnitude $r_{t,c} = \lVert \hat{\mathbf{p}}_{t,c} \rVert$. For the remainder of this section, we omit subscripts $t,c$ where the context is unambiguous.

\subsection{Geometric Decomposition of ACCDOA}
\label{subsect:Geometric_Decomp}

\begin{figure}[t]
    \centering
    \includegraphics[width=0.99\linewidth]{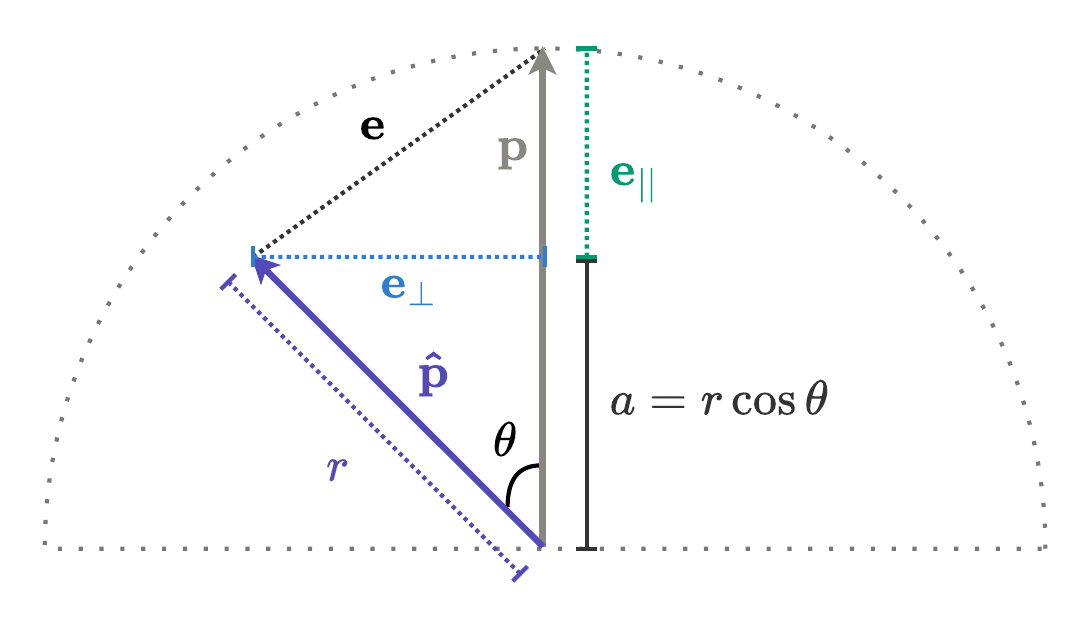}
    \caption{Geometric decomposition of the ACCDOA regression error ($\mathbf{e}$). The residual between the ground-truth target ($\mathbf{p}$, gray) and the model prediction ($\hat{\mathbf{p}}$, purple) is decoupled into orthogonal angular ($\mathbf{e}_\perp$, blue) and radial ($\mathbf{e}_\parallel$, green) components. To combat detection timidity, MAGENTA extracts the scalar projection $a$ of the prediction onto the target and uses the resulting radial under-prediction gap ($1 - a$) in an asymmetric penalty. This allows the network to decouple activity detection from spatial localization.}
    \label{fig:MAGENTA_Diag}
\end{figure}

The foundation of the MAGENTA framework is the geometric decomposition of the regression error for active frames. While the ACCDOA format unifies detection and localization into a single vector, standard regression losses, such as MSE, treat this vector as a generic point in Euclidean space. Consequently, they optimize the Cartesian coordinates $\{x,y,z\}$ independently, failing to distinguish between two fundamentally distinct types of physical errors: detection (magnitude) and localization (direction).

To resolve this ambiguity, we decompose the residual vector $\mathbf{e}=\mathbf{p} - \hat{\mathbf{p}}$ into two physically meaningful, orthogonal components:
\begin{align}
    \mathbf{e}_{\parallel} \, &= \langle \mathbf{e}, \mathbf{p} \rangle \, \mathbf{p} \, , \label{eqn:E_Parallel}\\
    \mathbf{e}_{\perp} &= \mathbf{e} - \mathbf{e}_{\parallel} \, . \label{eqn:E_Perp}
\end{align}

\noindent By orthogonality, $\lVert \mathbf{e} \rVert ^2 = \lVert \mathbf{e}_\parallel \rVert ^2 + \lVert \mathbf{e}_\perp \rVert ^2$. Here, the parallel component $(\mathbf{e}_{\parallel})$ represents the error along the axis of the true DOA. Because the vector's length dictates activity, this parallel error effectively measures the \textbf{event detection} accuracy in the correct direction. Conversely, the perpendicular component $(\mathbf{e}_{\perp})$ represents the orthogonal error, which specifically isolates the \textbf{spatial localization} accuracy.

To address the ``detection timidity'' characteristic of long-tailed learning, we utilize the scalar projection as a proxy for directional confidence:
\begin{equation}
    a = \langle \hat{\mathbf{p}}, \mathbf{p} \rangle = r\cos{\theta},
\end{equation}

\noindent where $\theta$ is the angular discrepancy between $\mathbf{p}$ and $\hat{\mathbf{p}}$. As illustrated in Fig.~\ref{fig:MAGENTA_Diag}, the scalar projection $a$ represents the predicted magnitude mathematically aligned with the correct ground-truth direction. To systematically combat ``detection timidity", we utilize this projection to construct an asymmetric, \textbf{under-prediction} radial penalty strictly for active frames:
\begin{equation}
    \label{equation:under_only_penalty}
    \mathcal{L}_\mathrm{under} = ([1 - a]_+)^2,
\end{equation}

\noindent where $[x]_+ = \max(0, x)$. Crucially, the geometric formulation is one-sided. If the model predicts a magnitude greater than 1.0 (over-confidence), it incurs zero penalty. Because strong detection confidence is highly desirable for rare events in heavily imbalanced scenarios, this asymmetric focus exclusively penalizes timidity. 

For spatial localization, the angular penalty must dictate how strictly the model is corrected for pointing in the wrong direction. To accommodate different acoustic complexities, we propose two variants based on the perpendicular error $\mathbf{e}_\perp$. 
\begin{enumerate}
    \item \textbf{Perpendicular Error Loss ($\mathcal{L}_\perp$)}: Penalizes the squared magnitude of the perpendicular error:
        \begin{equation}
            \label{eqn:L_perp}
            \mathcal{L}_\perp 
            = \lVert\mathbf{e}_{\perp}\rVert^2 
            = r^2\sin^2\theta 
            = r^2 - a^2 \,.
        \end{equation}

        This couples the angular penalty with the prediction magnitude $r$, meaning the model is penalized less for angular errors when it is uncertain.
        
    \item \textbf{Magnitude-Invariant Angular Loss ($\mathcal{L}_\mathrm{mia}$)}: To encourage rapid convergence of localization even for timid predictions, we decouple the angle from the magnitude:
        \begin{equation}
            \label{eqn:L_mia}
            \mathcal{L}_\mathrm{mia} = \sin^2\theta = 1 - \cos^2\theta \, .
        \end{equation}
\end{enumerate}

As analyzed later in Section~\ref{sec:Experimental_Results}, the choice between these two variants depends heavily on the acoustic complexity of the training corpus. Specifically, $\mathcal{L}_\mathrm{mia}$ enforces strict spatial gradients during severe data scarcity, while $\mathcal{L}_\perp$ acts as a confidence-weighted spatial regularizer suitable for synthetic data augmentation~\cite{roman2024spatial, yeow2025_enhancing}.

\subsection{Static Rarity-Aware Prior}
\label{subsect:Rarity_Prior}

In long-tailed datasets, naive uniform weighting allows frequent ``head" classes to dominate the gradient descent process~\cite{tang2020long}. While researchers typically counteract this using the proportional inverse of class frequency~\cite{cui2019class, imoto2022impact, zhang2024unified, cao2019LDAM}, direct scaling often overcompensates extreme tail classes, leading to gradient instability. To prevent head-class dominance while safely amplifying the faint optimization signals of rare events, we introduce a baseline compensation mechanism, shown in the Active Modulation (top) branch of Fig.~\ref{fig:magenta_flowchart}. 

We utilize a standard, static prior derived from the Effective Number of Samples (ENS)~\cite{cui2019class}, which accounts for data redundancy. Let $n_c$ be the total active frame count for class $c$, and $N = \sum_c n_c$. Using $\beta = (N-1)/N$, the unnormalized ENS-based rarity prior $\pi'_c$ for class $c$ is defined as the inverse of the effective sample size:
\begin{equation}
    \label{equation:Pi_C}
    \pi_c' = \frac{1-\beta}{1-\beta^{n_c}} \, .
\end{equation}

To maintain the scale of the original loss landscape and prevent gradient explosion~\cite{cao2019LDAM}, this prior is normalized to have a unit mean across all $C$ classes:
\begin{equation}
    \label{equation:Pi_C_Norm}
    \pi_c = \frac{\pi_c'}{\frac{1}{C} \sum_{j=1}^C \pi_j'} \, .
\end{equation}

\noindent The resulting normalized prior $\pi_c$ serves as a static boost factor for the radial loss:
\begin{equation}
    \mathcal{L}^\mathrm{stat}_{\mathrm{rad}, c} = (1 + \pi_c) \, \mathcal{L}_\mathrm{under}.
\end{equation}

Here, the additive term of $1$ acts as the base regression weight, preserving the standard unweighted penalty for all classes, while $\pi_c$ functions strictly as an additive margin. The base term is needed to prevent the vanishing of the loss weight for highly frequent head classes (where $\pi_c \approx 0$). Effectively, the combined multiplier forces the model to prioritize recall for rare events without sacrificing the baseline optimization of frequent events.

\subsection{Dynamic Detection Difficulty}
\label{subsect:Difficulty_Prior}

Relying solely on static statistical rarity for loss re-weighting can be suboptimal~\cite{park2021influence, ridnik2021asymmetric, ren2022balanced, Tan_EqualizationLosses_2023}, as statistical rarity does not strictly equate to learning difficulty~\cite{tian2022striking}. In complex acoustic environments, a rare event (e.g., \textit{Knock}) may be spectrally distinct and easy to learn, while a frequent event (e.g., \textit{Music}) may be highly diffuse.

To address this discrepancy, we propose an online difficulty estimation mechanism that continuously monitors the model's instantaneous performance. We define the instantaneous detection difficulty $\epsilon^{\mathrm{rad}}_{t,c}$ for an active frame at time $t$ as the magnitude of the under-prediction error:
\begin{equation}
\label{eqn:rad_diff}
\epsilon_{t,c}^{\mathrm{rad}} = [1 - a_{t,c}]_+ \, .
\end{equation}

To stabilize this metric, we average the instantaneous errors across all active frames for class $c$ within the current batch to obtain $\bar{\epsilon}^{\mathrm{rad}}_c[k]$. This batch-mean is then aggregated into a global, historical difficulty metric $\delta^{\mathrm{rad}}_c$ using an Exponential Moving Average (EMA) update with momentum $\alpha = 0.75$:
\begin{equation}
    \label{eqn:ema_update}
    \delta^{\mathrm{rad}}_c[k] = \alpha \, \delta^{\mathrm{rad}}_c[k-1] 
    + (1 - \alpha) \, \bar{\epsilon}^{\mathrm{rad}}_c[k] \, .
\end{equation}

This EMA formulation provides a stable, data-driven estimate of the model's struggle to detect a specific class. Similar to the static prior in~\eqref{equation:Pi_C_Norm}, this difficulty score is normalized to a unit mean across all $C$ classes to yield $\bar{\delta}_c^{\mathrm{rad}}$. This unit-mean normalization is strictly necessary to preserve the global scale of the loss landscape; because $\bar{\delta}_c^{\mathrm{rad}}$ is applied as a direct multiplier in the final radial penalty, failing to normalize it would cause the global gradients to arbitrarily drift as raw difficulties fluctuate. The normalized score acts as the dynamic multiplier for the adaptive radial penalty:
\begin{equation}
    \label{eqn:combined_boost_rad}
    \mathcal{L}^\mathrm{adapt}_{\mathrm{rad},c} = 
    (1 + \pi_c) \, 
    \bar{\delta}_c^{\mathrm{rad}} \, 
    \mathcal{L}_\mathrm{under} \, .
\end{equation}

\subsection{Adaptive Inactive Suppression}

For inactive, silent frames ($\lVert\mathbf{p}\rVert = 0$), standard regression objectives uniformly minimize the prediction norm $\lVert \hat{\mathbf{p}} \rVert^2$ to suppress false positives. However, rare classes inherently possess vastly more inactive frames than frequent classes~\cite{ridnik2021asymmetric, Tan_EqualizationLosses_2023, wang2021seesaw}. In a continuous SELD pipeline, an ``inactive frame'' for a specific class represents any time frame across the entire training corpus where that sound is not occurring. 

Let $n_\mathrm{total}$ represent the total number of frames in the dataset, and $n_c$ represent the active frames for class $c$. The total number of inactive frames for that class is strictly $n_\mathrm{total} - n_c$. Because a rare tail class possesses an exceptionally low $n_c$, it is heavily penalized as ``inactive'' for nearly the entire duration of the dataset. Under a uniform penalty, this overwhelming volume of silent frames creates gradient dominance~\cite{zhang2024unified,imoto2021impact, imoto2022impact}, systematically inducing ``detection timidity".

To counteract this, we initially scale the suppression penalty by a static inverse rarity prior, $w_c = (1 + \pi_c)^{-1}$. This explicitly lowers the mathematical risk threshold required for the model to predict tail classes during silence. This static weighted inactive loss becomes
\begin{equation}
\label{equation:Weighted_Inactive}
    \mathcal{L}^\mathrm{stat}_{\mathrm{inact}, c} = w_c \, \lVert \hat{\mathbf{p}} \rVert^2 \, .
\end{equation}

While effective at recovering recall, relying purely on this static inverse weighting introduces a precision-recall trade-off. Indiscriminately lowering the silence penalty for tail classes encourages the model to erroneously predict rare events during silent periods~\cite{tian2022striking}. To safely restore precision, we extend our dynamic difficulty framework to the inactive domain by tracking the energy of false positive predictions:
\begin{equation}
    \label{eqn:inactive_diff}
    \epsilon^\mathrm{inact}_{t,c} = \lVert \hat{\mathbf{p}}_{t,c} \rVert^2 \, .
\end{equation}

These false positive energies are aggregated into a global, class-wise metric $\delta^\mathrm{inact}_c$, using the EMA update described in~\eqref{eqn:ema_update}, and normalized to yield $\bar{\delta}_c^{\mathrm{inact}}$. A consistently high $\bar{\delta}_c^{\mathrm{inact}}$ indicates that the model systematically yields spurious detections for class $c$. We use this metric to directly modulate the static inverse prior, creating an adaptively weighted inactive loss:
\begin{equation} 
\label{equation:Weighted_Inactive_Dyn} 
\mathcal{L}^\mathrm{adapt}_{\mathrm{inact}, c} = w_c \, \bar{\delta}_c^{\mathrm{inact}} \, \lVert \hat{\mathbf{p}} \rVert^2 \, .
\end{equation}

\subsection{Intrinsic Difficulty-Driven Annealing}

While static rarity priors prevent head-class gradient dominance early in training, they can become detrimental as the network converges. Persistently applying a heavy rarity penalty to a mastered class unnecessarily inflates recall at the cost of precision. To resolve this, MAGENTA autonomously phases out statistical priors based on the model's real-time competence through an intrinsic difficulty-driven annealing mechanism.

We first define an intrinsic decay factor ($\gamma_c$) by normalizing the current class difficulty against the hardest class: 
\begin{equation}
\label{eqn:gamma_active}
\gamma_c^{\mathrm{rad}} = \frac{\bar{\delta}_c^{\mathrm{rad}}}{\max_j(\bar{\delta}_j^{\mathrm{rad}})} \, .
\end{equation}

The static rarity prior is scaled by this factor, defining the annealed rarity prior $\pi^*_c \, = \, \gamma_c^{\mathrm{rad}} \, \pi_c$ as a self-pacing penalty. Early in training, a difficult tail class yields $\gamma_c^{\mathrm{rad}} \approx 1.0$, retaining the full benefit of the static prior. As the model learns the class and the under-projection gap shrinks, $\gamma_c^{\mathrm{rad}} \rightarrow 0$. This smooth decay autonomously removes the static rarity prior, restoring precision by reducing the risk of late-stage false positives.

The final adaptive radial penalty is constructed by applying this annealed rarity prior, combined with a direct linear scaling of the normalized dynamic difficulty $\bar{\delta}_c^{\mathrm{rad}}$: 
\begin{equation}
\label{eqn:final_active_loss}
\mathcal{L}^\mathrm{final}_{\mathrm{rad}, c} = (1 + \pi^*_c) \, \bar{\delta}_c^{\mathrm{rad}} \, \mathcal{L}_{\mathrm{under}} \, .
\end{equation}

This self-correcting logic is mirrored in the inactive suppression domain. We define a similar inactive decay factor $\gamma_c^{\mathrm{inact}}$ using the EMA of the false positive energy ($\bar{\delta}_c^{\mathrm{inact}}$):
\begin{equation}
    \label{eqn:gamma_c_inact}
    \gamma_c^{\mathrm{inact}} = \frac{\bar{\delta}_c^{\mathrm{inact}}}{\max_j(\bar{\delta}_j^{\mathrm{inact}})} \, .
\end{equation}

\noindent This factor recalculates the annealed inverse prior $w_c^*$ on the fly: 
\begin{equation}
\label{eqn:dynamic_w_c}
w_c^* = \frac{1}{1 + \gamma_c^{\mathrm{inact}} \, \pi_c} \, .
\end{equation}

\noindent The final adaptively weighted inactive loss therefore becomes:
\begin{equation}
\label{eqn:final_inactive_loss}
\mathcal{L}^\mathrm{final}_{\mathrm{inact}, c} = w_c^* \, \bar{\delta}_c^{\mathrm{inact}} \, \lVert \hat{\mathbf{p}} \rVert^2 \, .
\end{equation}

Through this mechanism, MAGENTA establishes a fully dynamic curriculum. It lowers the risk threshold to encourage tail-class predictions, but applies a strict suppression penalty when those predictions develop into systematic false positives.

\subsection{Unified Loss}
\label{subsect:Unified_Loss}

Ultimately, as shown at the rightmost nodes of Fig.~\ref{fig:magenta_flowchart}, the MAGENTA framework operates as a dynamic, dual-state objective. When a sound is actively occurring, the network minimizes the intrinsically annealed radial penalty with the chosen angular loss:
\begin{equation}
    \label{eqn:Combined_Active_Loss}
    \mathcal{L}_{\mathrm{act}, c} = 
    \mathcal{L}^\mathrm{final}_{\mathrm{rad}, c} + 
    \mathcal{L}_{\mathrm{ang}},
\end{equation}

\noindent where $\mathcal{L}_{\mathrm{ang}} \in \{\mathcal{L}_\perp, \mathcal{L}_{\mathrm{mia}}\}$, depending on the selected configuration. We define the unified loss, $\mathcal{L}_{\mathrm{total}}$, as the masked average over all time frames $T$ and classes $C$:
\begin{equation}
    \mathcal{L}_{\mathrm{total}} = \frac{1}{T C} \sum_{t,c}
    \begin{cases}
        \mathcal{L}_{\mathrm{act},c}, & \mathrm{if} \, \mathrm{active},\\
        \mathcal{L}^\mathrm{final}_{\mathrm{inact},c}, & \mathrm{otherwise}.
    \end{cases}
\end{equation}

Overall, MAGENTA ensures that gradient updates are physically interpretable and dynamically tuned. By decoupling detection, localization, and suppression, the model can learn to be aggressive in detecting difficult, rare events while maintaining high precision in localization and silence.

\section{Implementation Details}

\subsection{Dataset and Features}

We evaluate our proposed method on the STARSS23 dataset~\cite{shimada2023starss23}, the official corpus for DCASE 2023 Challenge Task 3. As one of the few publicly available SELD datasets featuring real-world environments, STARSS23 serves as a critical testbed for evaluating real-world performance, presenting a class imbalance factor exceeding $500\times$. However, it is limited in size due to the high cost of manual spatiotemporal annotation. Therefore, to improve model generalization while preserving this inherent long-tailed distribution, we employ the rotation of First-Order Ambisonics (FOA) signals~\cite{mazzon2019first, wang2023four}, effectively expanding the training set by a factor of eight. 

For feature extraction, we utilize the four-channel FOA format sampled at 24\,kHz. We compute log-Mel spectrograms concatenated with normalized intensity vectors~\cite{perotin2018crnn} to capture both spectral and spatial cues. The Short-Time Fourier Transform (STFT) is calculated using a 1024-point FFT with a Hann window of 1024 samples and a hop length of 480 samples (20 ms), and 64 Mel-frequency bands are used. 

In addition to training exclusively on real-world recordings, we conduct supplementary experiments utilizing the official DCASE synthetic dataset (DCASE-Syn) provided alongside the challenge~\footnote{\url{https://zenodo.org/records/6406873}}, which is used to supplement the STARSS23 training corpus. Incorporating DCASE-Syn allows us to validate the robustness of the MAGENTA framework within a hybrid training paradigm, where synthetic data is used to mitigate the scarcity of tail classes~\cite{zhang2023deep, yang2022survey}, yet inherent class imbalance and synthetic-to-real domain shift persist. 

\subsection{Model Training}

\begin{figure}[t]
    \centering
    \includegraphics[width=0.7\linewidth]{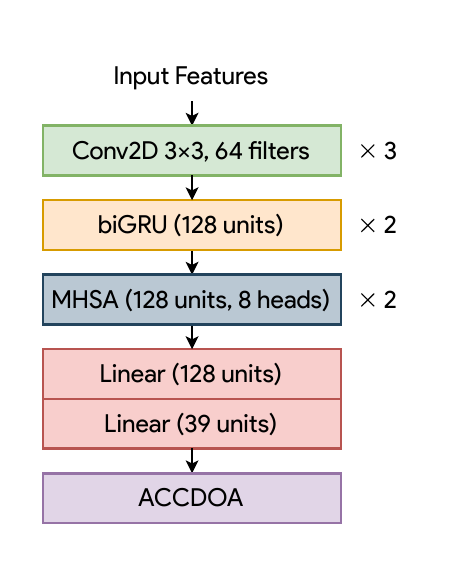}
    \caption{The SELDNet architecture~\cite{adavanne2018sound} used as the baseline model. The model follows a CRNN structure, mapping to the unified ACCDOA output space.}
    \label{fig:seldnet_block_diag}
\end{figure}

To isolate the impact of the proposed MAGENTA loss framework, we adopt the standard SELDNet architecture~\cite{adavanne2018sound, shimada2023starss23} as our backbone, as illustrated in Fig.~\ref{fig:seldnet_block_diag}. The model comprises three convolutional blocks with 64 filters each, followed by two bidirectional Gated Recurrent Unit (biGRU) layers and two Multi-Head Self-Attention (MHSA) blocks to capture global context and temporal dependencies. The final layers consist of fully connected units that map the latent features to the single-track ACCDOA representation.

All models are trained for 100 epochs using the Adam optimizer with a weight decay of $1 \times 10^{-4}$. We employ a OneCycle learning rate scheduler~\cite{smith2019super} with a peak learning rate of $1 \times 10^{-3}$ to ensure stable convergence. The batch size is set to 64.

To facilitate reproducibility and provide a clear overview of the MAGENTA framework's configuration, Table~\ref{tab:hyperparameters} summarizes the primary hyper-parameters and architectural design choices required for implementation.

\begin{table}[htbp]
\centering
\caption{Summary of MAGENTA Hyper-parameters and Design Choices}
\label{tab:hyperparameters}

\begin{tabularx}{\linewidth}{l c L}
\toprule
\textbf{Parameter} & \textbf{Recommended} & \textbf{Description} \\

\midrule
EMA Momentum ($\alpha$) & 0.75 & Controls the update rate of the intrinsic difficulty tracking. A value of 0.75 ensures stability against batch-level variance without becoming too sluggish (analyzed in Section~\ref{subsection:sensitivty_difficulty_momentum}). \\

\midrule
Angular Loss ($\mathcal{L}_\mathrm{ang}$) & $\mathcal{L}_\mathrm{mia}$ or $\mathcal{L}_\perp$ & $\mathcal{L}_\mathrm{mia}$ is recommended for strictly real-world data to force spatial learning. $\mathcal{L}_\perp$ is recommended when using synthetic data augmentation, acting as a spatial regularizer to prevent domain overfitting (analyzed in Section~\ref{subsection:robustness_under_synthetic}). \\

\midrule
Activity Threshold & 0.5 & The standard threshold in the ACCDOA representation~\cite{shimada2021accdoa} used to determine if a predicted vector corresponds to an active or inactive event. \\

\bottomrule
\end{tabularx}

\end{table}

\subsection{Evaluation Metrics}
\label{section:seld_eval_metrics}

Performance is assessed using the official SELD metrics from the DCASE 2023 Challenge Task 3~\cite{politis2020overview}. To accurately reflect the spatiotemporal coupling required in real-world environments, modern SELD evaluation relies on joint metrics that concurrently measure classification and localization accuracy. 

These joint evaluation metrics are divided into two complementary categories. The first category consists of location-dependent detection metrics, which assess detection performance while enforcing a strict spatial penalty. A predicted sound event is only considered a true positive if its predicted class matches the ground truth and its estimated spatial location falls within a specific angular distance threshold from the reference location. As per standard practice~\cite{politis2020overview, shimada2023starss23}, we apply a $20^\circ$ spatial threshold to compute the location-dependent error rate (\ER) and F-score (\Fone). 

The second category evaluates spatial precision. Crucially, these metrics only calculate errors between correctly associated class pairs. These include the class-dependent localization error (\LE), which measures the mean angular error of true positives, and the localization recall (\LR), which acts as a location-agnostic true positive rate. Notably, to account for undetected events, if a class receives zero recall, it incurs a maximum spatial penalty of $180^\circ$ for its individual \LE. Furthermore, all metrics aside from \ER\ are macro-averaged across all sound event classes. Consequently, complete detection failures on rare tail classes can disproportionately degrade the aggregated \LE\ performance.

Finally, to provide a single, unified benchmark for overall model comparison, we utilize the official aggregated SELD error (\ESELD) established by the DCASE Challenge organizers~\cite{politis2020overview}. In this formulation, \LE\ is divided by $180^\circ$ to normalize the spatial penalty to a $[0, 1]$ range, allowing it to be equally weighted alongside the other normalized metrics:
\begin{equation}
    \label{eqn:seld_error}
    \mathcal{E}_\mathrm{SELD} = 
    \frac{
    \mathrm{ER}_{\le20^\circ} + 
    (1 - \mathrm{F}_{\le20^\circ}) +
    \dfrac{\mathrm{LE}_\mathrm{CD}}{180^\circ} +
    (1 - \mathrm{LR}_\mathrm{CD})   
    }
    {4}.
\end{equation}

\noindent An effective SELD system should have low $\mathrm{ER}_{\le20^\circ}$, $\mathrm{LE}_\mathrm{CD}$, and \ESELD, as well as high $\mathrm{F}_{\le20^\circ}$ and $\mathrm{LR}_\mathrm{CD}$.

\section{Results and Discussion}
\label{sec:Experimental_Results}

This section presents experimental results evaluated on the validation set of the real-world STARSS23 dataset~\cite{shimada2023starss23}. All reported results are averaged over five runs. 

\subsection{Incremental Contribution of the MAGENTA Framework}
\label{subsection:incremental_contribution}

\begin{table*}[t]
\centering
\caption{Analysis of the incremental contribution of each component in the proposed MAGENTA framework. A (\checkmark) indicates an enabled component, while (-) indicates the baseline counterpart.}
\label{tab:ablation_configs}
\definecolor{LightGray}{gray}{0.9}  % Define a light gray color for the row
\begin{tabularx}{\textwidth}{l | l | ccc | cc | CCCCC}

\toprule

& & \multicolumn{3}{c|}{\textbf{Active Frames Loss}} & \multicolumn{2}{c|}{\textbf{Inactive Frames Loss}} & \multicolumn{5}{c}{\textbf{SELD Metrics}}  \\

\cmidrule(lr){3-5} \cmidrule(lr){6-7} \cmidrule(lr){8-12}

\textbf{ID} & \textbf{Description} & 
\textbf{Static} & \textbf{Dynamic} & \textbf{MIA} & 
\textbf{Static} & \textbf{Dynamic} 
& \ER $\downarrow$ & \Fone $\uparrow$ & \LE $\downarrow$ & \LR $\uparrow$ & \ESELD $\downarrow$ \\

\midrule
\multicolumn{12}{l}{\textit{Baselines}} \\
\midrule

\textbf{B1} & Standard MSE & - & - & - & - & -
& 0.616 & 22.0\% & $60.3^\circ$ & 31.2\% & 0.605 \\

\textbf{B2} & Geom. Decomp. & - & - & - & - & -
& 0.622 & 23.3\% & $63.9^\circ$ & 32.4\% & 0.605 \\

\midrule
\multicolumn{12}{l}{\textit{MAGENTA Framework}} \\
\midrule

\textbf{M1} & Static Active & \checkmark & - & - & - & - 
& 0.634 & 29.0\% & $24.7^\circ$ & 43.2\% & 0.512 \\

\textbf{M2} & + Dyn Active & \checkmark & \checkmark & - & - & - 
& 0.620 & 29.8\% & $22.7^\circ$ & 45.6\% & 0.498 \\

\textbf{M3} & + MIA & \checkmark & \checkmark & \checkmark & - & - 
& 0.625 & 30.6\% & $20.6^\circ$ & 45.7\% & 0.494 \\

\textbf{M4} & + Static Inactive & \checkmark & \checkmark & \checkmark & \checkmark & -
& 0.627 & 30.2\% & $21.3^\circ$ & 49.6\% & 0.487 \\

\textbf{M5} & + Dyn Inactive & \checkmark & \checkmark & \checkmark & \checkmark & \checkmark
& 0.621 & 30.5\% & $\textbf{20.5}^\circ$ & \textbf{49.8}\% & 0.483 \\

\rowcolor{LightGray}
\textbf{D1} & M5 w/ Intrinsic Decay & \checkmark & \checkmark & \checkmark & \checkmark & \checkmark
& \textbf{0.612} & \textbf{31.0}\% & $20.7^\circ$ & 49.5\% & \textbf{0.481} \\

\bottomrule
\end{tabularx}
\raggedright
\footnotesize{\textit{Note: Static Active ($\pi_c$) refers to \eqref{equation:Pi_C_Norm}; Dynamic Active ($\bar{\delta}_c^{\mathrm{rad}}$) to \eqref{eqn:combined_boost_rad}; MIA ($\mathcal{L}_{\mathrm{mia}}$) to \eqref{eqn:L_mia}; Static Inactive ($w_c$) to \eqref{equation:Weighted_Inactive}; Dynamic Inactive ($\bar{\delta}_c^{\mathrm{inact}}$) to \eqref{equation:Weighted_Inactive_Dyn}.}}
\end{table*}

Table~\ref{tab:ablation_configs} details the incremental contributions of each component within the proposed MAGENTA framework. The standard MSE baseline (B1) demonstrates the inherent limitations of standard regression objectives on heavily imbalanced data. Because MSE is dominated by the overwhelming number of inactive frames, it heavily suppresses predictions to avoid gradient penalties, yielding severe ``detection timidity" and a poor \LR\ of 31.2\%. Consequently, the MSE baseline completely fails to detect several tail classes. Under the DCASE macro-averaging evaluation framework, these complete detection failures incur maximum $180^\circ$ penalties, which drastically skew the baseline's \LE\ to an inflated $60.3^\circ$. 

Introducing the static rarity prior (M1) penalizes under-predictions on rare classes. By recovering predictions for these missing tail classes, M1 replaces those maximum $180^\circ$ penalties with actual measured angular errors, sharply improving \Fone (29.0\%), \LR\ (43.2\%), and dropping \LE\ to $24.7^\circ$. Given that complete detection failures can substantially increase the averaged \LE, the improvement in \LE\ should therefore be interpreted together with \Fone\ and \LR, which directly reflect recovery of previously missed tail events.

Incorporating the dynamically tracked active prior (M2) further refines these gains, dropping the overall \ESELD\ to 0.498. Next, the integration of the MIA loss (M3) decouples the angular penalty from the prediction magnitude. By forcing the model to learn robust spatial representations even when detection confidence is low, \LE\ decreases to $20.6^\circ$ and \Fone\ improves to 30.6\%.

Conversely, applying a static inverse rarity prior to inactive frames (M4) reveals the danger of standard LTL techniques during silence. While M4 deliberately lowers the risk threshold for predicting rare events (improving \LR\ to 49.6\%), it introduces a natural precision-recall trade-off~\cite{tian2022striking}. The degradation in \Fone\ and \LE\ confirms that aggressively scaling down silence penalties for rare classes inadvertently teaches the network to over-predict them as false positives. The dynamic inactive difficulty metric (M5) effectively mitigates this specific over-prediction problem. By adaptively tracking and penalizing false positive energy during silent periods, M5 preserves a robust \LR\ of 49.8\% while recovering precision.

Finally, the intrinsic difficulty-driven annealing mechanism (D1) completes the self-correcting MAGENTA architecture, achieving the best \ESELD\ of 0.481. Crucially, autonomously decaying the static rarity priors as the network masters specific classes maximizes overall precision, evidenced by the lowest \ER\ (0.612) and best \Fone\ (31.0\%).

\subsection{Analysis of Class-Wise Performance}
\label{subsect:classwise_longtail}

\begin{figure*}[t]
    \centering
    \includegraphics[width=0.99\linewidth]{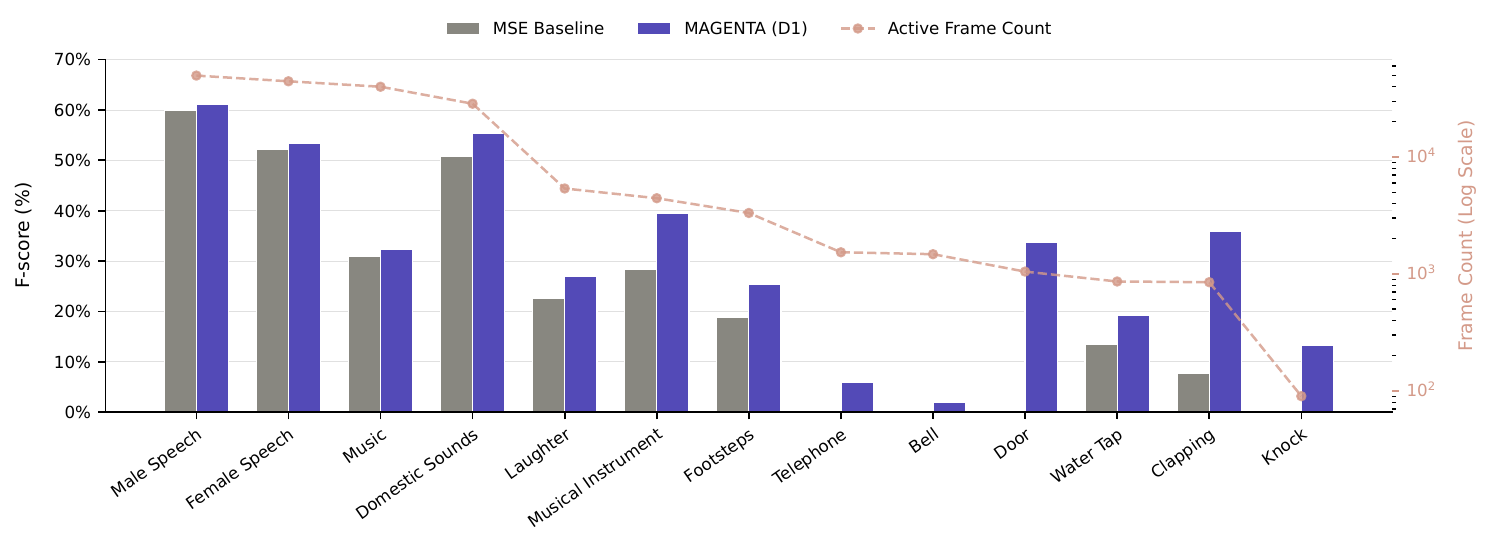} 
    \caption{Class-wise F-score (\Fone) evaluated on the STARSS23 validation set, sorted descending by active frame count (dashed line). The MSE baseline exhibits ``detection timidity" in the tail, whereas MAGENTA (D1) successfully recovers performance for rare events without degrading head class precision.}
    \label{fig:classwise_fscore}
\end{figure*}

\begin{figure*}[t]
    \centering
    \includegraphics[width=0.99\linewidth]{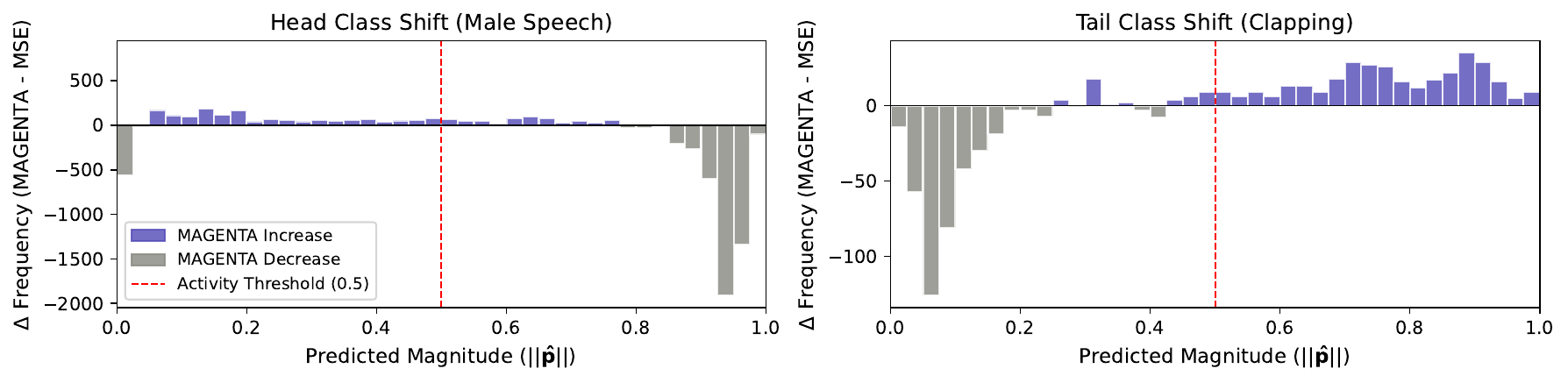}
    \caption{Comparison of predicted ACCDOA magnitude ($||\hat{\mathbf{p}}||$) shifts for active frames, contrasting a head class (\textit{Male Speech}) against a tail class (\textit{Clapping}). \textit{Clapping} is selected to represent the tail distribution here, rather than the absolute rarest class (\textit{Knock}), to provide sufficient validation sample density for a representative visualization of the behavioral shift. The plots display the net difference in prediction distributions ($\Delta$ Frequency) per discrete magnitude bin. Negative grey bars denote predictions suppressed by MAGENTA relative to the MSE baseline, while positive purple bars denote an increase in predictions. In the ACCDOA format, a magnitude exceeding 0.5 denotes an active event. For the head class (left), MAGENTA safely maintains confident predictions above the threshold, while the standard MSE baseline suffers from severe ``detection timidity" in the tail class (right). MAGENTA successfully recovers performance by eliminating these near-zero predictions and shifting a significant portion of tail predictions across the 0.5 threshold, substantially improving recall while preserving precision.}
    \label{fig:magnitude_histogram}
\end{figure*}

To better understand these gains, Fig.~\ref{fig:classwise_fscore} illustrates the class-wise \Fone\ of the intrinsically decayed MAGENTA configuration (D1) against the standard MSE baseline, overlaid with the logarithmic active frame count.

The visualization illustrates the ``detection timidity'' problem inherent in standard regression objectives. While the MSE baseline achieves respectable recall on data-rich head classes (e.g., \textit{Male Speech}, \textit{Female Speech}), its performance degrades substantially as data scarcity increases. For the most extreme ``tail" classes (e.g., \textit{Telephone}, \textit{Door}, and \textit{Knock}), the MSE model yields near-zero \Fone. To avoid the massive gradient penalties accumulated during silent frames, the standard MSE model unconditionally suppresses its predictions.

To investigate the physical mechanics driving this failure, we examine the net shift in prediction frequencies ($\Delta$ Frequency) for ground-truth active frames between MAGENTA and the MSE baseline. Mathematically, the net shift for a given magnitude bin $b$ is computed as:
\begin{equation}
    \Delta f(b) = N_{\mathrm{MAGENTA}}(r \in b) - N_{\mathrm{MSE}}(r \in b)
\end{equation}

\noindent where $N(r \in b)$ denotes the total number of active frames where the predicted magnitude $r$ falls within the boundaries of bin $b$. As shown in Fig.~\ref{fig:magnitude_histogram} (left), the shift for a data-rich head class (\textit{Male Speech}) is relatively stable.

MAGENTA slightly regularizes over-confident predictions near 1.0, while safely maintaining the overall prediction mass well above the 0.5 activity threshold. In contrast, the behavioral shift for the tail class (\textit{Clapping}) in Fig.~\ref{fig:magnitude_histogram} (right) is distinct. The pronounced shift near a magnitude of 0.0 indicates that the standard MSE baseline collapses its predictions to near-zero. This confirms that under extreme imbalance, standard regression inherently suppresses the prediction vector magnitude to minimize global error, inducing ``detection timidity". 

By geometrically decoupling the regression error and applying dynamically weighted modulation, MAGENTA actively resolves this timidity. It eliminates this near-zero collapse and successfully shifts a significant proportion of tail class predictions across the 0.5 threshold. Consequently, MAGENTA achieves an \Fone\ of 33.7\% and 13.3\% for the \textit{Door} and \textit{Knock} classes, respectively, where the baseline failed completely. Crucially, the dynamic difficulty-driven annealing mechanism ensures this tail recovery does not degrade head-class precision, highlighting the framework's robustness in mitigating extreme acoustic class imbalance.

\subsection{Analysis of Learning Dynamics}
\label{subsect:learning_dynamics}

\begin{figure}[t]
    \centering
    \includegraphics[width=0.99\linewidth]{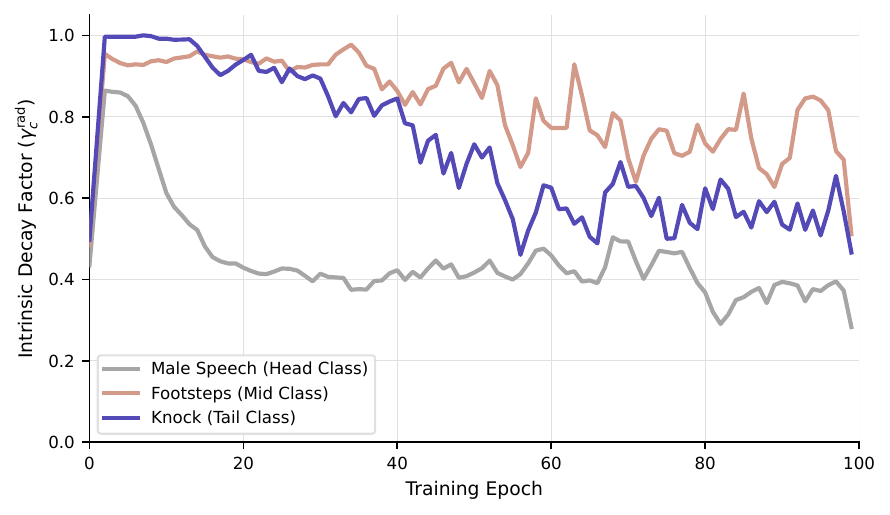} 
    \caption{Trajectories of the intrinsic decay factor ($\gamma_c^{\mathrm{rad}}$) during training. MAGENTA autonomously curates a unique learning pace per class based on instantaneous difficulty, rather than static frame counts. Notably, it sustains a high penalty for persistently difficult mid-classes (\textit{Footsteps}) while smoothly annealing for tail classes (\textit{Knock}) once they are learned.}
    \label{fig:learning_dynamics}
\end{figure}

To illustrate how MAGENTA mitigates the precision-recall trade-off~\cite{tian2022striking}, Fig.~\ref{fig:learning_dynamics} tracks the active intrinsic decay factor ($\gamma_c^{\mathrm{rad}}$) during training for three representative sound classes: a data-rich head class (\textit{Male Speech}), a moderate middle class (\textit{Footsteps}), and a scarce tail class (\textit{Knock}).

This visualization highlights a critical advantage of MAGENTA's dynamic difficulty tracking. As expected, common and acoustically distinct head classes such as \textit{Male Speech} are mastered early in training. Consequently, its corresponding $\gamma_c^{\mathrm{rad}}$ rapidly decays, autonomously phasing out the static rarity prior to prevent late-stage over-prediction and preserve high precision. Conversely, the extremely rare \textit{Knock} class requires greater gradient focus. For this class, $\gamma_c^{\mathrm{rad}}$ remains near 1.0 for the first 30 epochs, maintaining a heavy rarity penalty until the network successfully learns its acoustic signature, after which the penalty smoothly anneals.

Crucially, the trajectory of the \textit{Footsteps} class validates our premise that statistical rarity does not strictly equate to learning difficulty. Despite possessing substantially more training frames than \textit{Knock}, footsteps are acoustically broadband, diffuse, and easily masked by ambient noise, presenting a persistent detection challenge. The difficulty-driven curriculum of MAGENTA dynamically adapts to this, sustaining a high rarity penalty for \textit{Footsteps} deep into the training cycle. By dynamically linking the curriculum to real-time spatial and semantic competence, MAGENTA optimizes learning for the unique acoustic reality of each class.

\subsection{Ablation Studies on Core Design Choices}
\label{subsect:core_design_choices}

\begin{table}[t]
\centering
\caption{Ablation study evaluating the core design choices of the MAGENTA framework.}
\label{tab:Design_Choices}
\definecolor{LightGray}{gray}{0.9}  % Define a light gray color for the row
\begin{tabularx}{\linewidth}{c | l | CCCCC}

\toprule
ID & Description & \ER $\downarrow$ & \Fone $\uparrow$ & \LE $\downarrow$ & \LR $\uparrow$ & \ESELD $\downarrow$ \\

\midrule
\multicolumn{7}{l}{\textit{Necessity of Decomposition}} \\
\midrule

\rowcolor{LightGray}
\textbf{M5} & MAGENTA
& 0.621 & 30.5\% & $20.5^\circ$ & 49.8\% & 0.483 \\ 

\textbf{N1} & M5 - Decomp
& 0.625 & 29.2\% & $21.4^\circ$ & 48.0\% & 0.493 \\

\midrule
\multicolumn{7}{l}{\textit{Static vs. Dynamic Modulation}} \\
\midrule

\rowcolor{LightGray}
\textbf{M5} & MAGENTA
& 0.621 & 30.5\% & $20.5^\circ$ & 49.8\% & 0.483 \\ 

\textbf{R1} & Rarity-Only
& 0.640 & 29.4\% & $20.5^\circ$ & 48.2\% & 0.494 \\

\midrule
\multicolumn{7}{l}{\textit{Angular Loss Formulation}} \\
\midrule

\rowcolor{LightGray}
\textbf{D1} & M5 + Decay 
& 0.612 & 31.0\% & $20.7^\circ$ & 49.5\% & 0.481 \\

\textbf{D2} & D1 w/ $\mathcal{L}_\perp$ 
& 0.615 & 30.5\% & $21.9^\circ$ & 48.4\% & 0.487 \\

\bottomrule
\end{tabularx}
\end{table}

Table~\ref{tab:Design_Choices} isolates the specific impacts of the core mechanics of MAGENTA, evaluating geometric decomposition, dynamic difficulty tracking, and angular loss formulation.

First, we examine the necessity of geometrically treating the ACCDOA vector. We compare the dynamically tracked MAGENTA framework (M5) against a non-decomposed variant (N1), where the identical static rarity priors and dynamic difficulty metrics are applied directly to modulate the standard MSE objective. Removing the geometric decomposition results in a distinct performance degradation, most notably worsening spatial precision with \LE\ increasing to $21.4^\circ$. This degradation exposes the mathematical conflict inherent in scaling Cartesian distances for long-tailed SELD. Indiscriminately scaling both activity and directional errors simultaneously leads to unstable gradients. The superior performance of M5 confirms that physically separating the regression error into orthogonal components is a prerequisite for robust, targeted modulation.

Building on this decoupled space, we next validate the dynamic tracking mechanism by contrasting M5 against a static-only baseline (R1) that omits the dynamic difficulty modulation. The rarity-only baseline suffers from an increased \ER\ of 0.640. Because static statistical weights operate on the flawed assumption that rarity strictly equates to difficulty, they can inadvertently force the network to over-predict rare classes, increasing false positive rates. In contrast, the dynamic difficulty tracking of M5 acts as a critical self-correcting mechanism, automatically dampening the impact of priors as the model grows competent to robustly improve recall while preserving precision.

Finally, we evaluate the spatial penalty by comparing the MIA formulation (D1) against the perpendicular error loss (D2). In heavily imbalanced datasets, standard regression suppresses prediction magnitudes for tail classes to near-zero. By mathematically coupling the angular penalty to this suppressed magnitude, $(\mathcal{L}_\perp)$ can artificially delay spatial learning. In contrast, the D1 configuration utilizing the MIA loss lowers \LE\ to $20.7^\circ$ and achieves the best overall \ESELD\ of 0.481. By enforcing spatial gradients regardless of the model's detection confidence, $\mathcal{L}_\mathrm{mia}$ ensures robust localization even under severe detection timidity.

\subsection{Comparison with Classification-based Loss-Balancing}
\label{subsect:classification}

\begin{table}[t]
\centering\caption{Performance comparison between two-branch architectures (utilizing classification-based LTL) and the unified ACCDOA architecture.}
\label{tab:two_branch_comparison}
\definecolor{LightGray}{gray}{0.9}  % Define a light gray color for the row

\begin{tabularx}{\linewidth}{CC CCCCC}

\toprule

\textbf{SED} & \textbf{DOA} & \ER $\downarrow$ & \Fone $\uparrow$ & \LE $\downarrow$ & \LR $\uparrow$ & \ESELD $\downarrow$ \\

\midrule
\multicolumn{7}{l}{\textit{Disjointed Multi-Task (Two-Branch) --- 745k Params}} \\

\midrule
BCE & MSE 
& 0.700 & 25.4\% & $25.8^\circ$ & 44.8\% & 0.535 \\

Focal~\cite{lin2017focal} & MSE 
& 0.658 & 28.0\% & $23.8^\circ$ & 44.2\% & 0.517 \\

IFL~\cite{zhang2024unified} & MSE 
& 0.680 & 29.2\% & $21.7^\circ$ & 44.1\% & 0.517 \\

\midrule 
\multicolumn{7}{l}{\textit{Unified Regression (ACCDOA Single-Branch) --- 727k Params}} \\
\midrule
\multicolumn{2}{c}{MSE} 
& 0.616 & 22.0\% & $60.3^\circ$ & 31.2\% & 0.605 \\

\rowcolor{LightGray}
\multicolumn{2}{c}{MAGENTA (D1)} 
& \textbf{0.612} & \textbf{31.0}\% & $\textbf{20.7}^\circ$ & \textbf{49.5}\% & \textbf{0.481} \\

\bottomrule
\end{tabularx}
\end{table}

To validate the efficacy of maintaining a unified regression space, we compare the proposed MAGENTA framework against traditional multi-task learning techniques. Established LTL methods, such as Focal Loss~\cite{lin2017focal} and Inverse Frequency Loss (IFL), operate strictly on categorical probabilities. Consequently, they cannot be directly applied to the continuous ACCDOA vector. Leveraging these standard LTL techniques requires reverting to a disjointed, two-branch architecture with separate task heads for SED and DOA estimation~\cite{adavanne2018sound}. 

As demonstrated in Table~\ref{tab:two_branch_comparison}, the standard two-branch baseline optimized using binary cross entropy (BCE) and MSE achieves an \ESELD\ of 0.535, noticeably outperforming the standard ACCDOA baseline (0.605). To ensure a fair comparison, we conducted a brief hyperparameter grid search to tune the trade-off between the classification and regression heads, confirming that an equal weight distribution yielded optimal performance~\cite{phan2020multitask, adavanne2018sound}. However, tuning a global scalar parameter is theoretically limited in addressing long-tailed distributions. Because a global weight scales the loss uniformly across all classes, it cannot simultaneously penalize frequent head classes while lowering the risk threshold for rare tail classes. Thus, it is mathematically incapable of resolving class-specific detection timidity.

Applying categorical LTL losses to the SED branch further improves detection, increasing \Fone\ from 25.4\% to 28.0\% with Focal Loss~\cite{lin2017focal} and 29.2\% with IFL~\cite{zhang2024unified}. Both methods reduce \ESELD\ to 0.517, demonstrating that classification-based reweighting can partially mitigate long-tailed detection errors. Nevertheless, these improvements remain limited by the two-branch formulation. The task-balancing weight is a single global scalar, so it adjusts the overall trade-off between SED and DOA estimation uniformly across all classes. It cannot distinguish frequent head classes from rare tail classes, nor can it adapt to class-specific learning difficulty during training.

MAGENTA addresses this limitation directly inside the continuous ACCDOA space. By geometrically decomposing the regression error and applying class-specific dynamic modulation, MAGENTA (D1) achieves the best overall result, reducing \ESELD\ to 0.481. It also obtains the highest \Fone\ of 31.0\% and \LR\ of 49.5\%, while using slightly fewer parameters than the two-branch models (727k vs. 745k). These results show that MAGENTA retains the efficiency of unified ACCDOA regression while integrating selective emphasis on under-learned classes.

\subsection{Sensitivity to Difficulty Momentum}
\label{subsection:sensitivty_difficulty_momentum}

\begin{table}[t]
\centering
\caption{Impact of the EMA momentum factor $(\alpha)$ on the performance of the intrinsically decayed MAGENTA configuration (D1).}
\label{tab:vary_ema}
\definecolor{LightGray}{gray}{0.9}  % Define a light gray color for the row
\begin{tabular}{c ccccc}

\toprule
\textbf{Alpha} ($\alpha$) & \ER $\downarrow$ & \Fone $\uparrow$ & \LE $\downarrow$ & \LR $\uparrow$ & \ESELD $\downarrow$ \\

\midrule

0.01
& 0.623 & 30.6\% & $\textbf{20.0}^\circ$ & 49.0\% & 0.485 \\

0.25
& 0.622 & 30.3\% & $20.4^\circ$ & 48.8\% & 0.486 \\

0.50 
& 0.618 & 30.3\% & $20.8^\circ$ & 48.6\% & 0.486 \\

\rowcolor{LightGray}
0.75
& \textbf{0.612} & \textbf{31.0}\% & $20.7^\circ$ & \textbf{49.5}\% & \textbf{0.481} \\

0.99
& 0.616 & 30.8\% & $20.6^\circ$ & 48.8\% & 0.484 \\

\bottomrule
\end{tabular}
\end{table}

Table~\ref{tab:vary_ema} details the performance of the intrinsically decayed MAGENTA configuration (D1) across varying levels of the EMA momentum factor $\alpha$. This parameter dictates the update rate of the intrinsic difficulty tracking, controlling how rapidly the self-paced curriculum adapts to the network's real-time performance. Optimal balance is achieved using $\alpha = 0.75$, yielding the lowest \ESELD\ of 0.481 and the highest \Fone\ of 31.0\%. This indicates that a moderately stable historical memory is needed to reliably estimate class-wise difficulty without being too impacted by batch-level variance. 

Deviating from this optimum value degrades performance. Significantly lowering the momentum $(\alpha \leq 0.50)$ makes the difficulty tracking overly sensitive to the composition of individual mini-batches. While an extremely reactive tracker $(\alpha = 0.01)$ marginally improves \LE\ to $20.0^\circ$, it generates a noisy curriculum that degrades spatial precision, as evidenced by a drop in \LR\ and an overall worse \ESELD. Conversely, an aggressively high momentum $(\alpha = 0.99)$ makes the tracking too sluggish. By failing to decay the rarity priors once a class is mastered, the network is forced to maintain unnecessary gradient penalties, leading to increased false positive rates and \ER.

\subsection{Robustness Under Synthetic Data Augmentation}
\label{subsection:robustness_under_synthetic}

\begin{table}[t]
\centering
\caption{Performance of the MAGENTA framework under real-world only and real+synthetic training conditions.}
\label{tab:sota_dataset}
\begin{tabularx}{\linewidth}{l CCCCC}

\toprule
\textbf{ID} & \ER $\downarrow$ & \Fone $\uparrow$ & \LE $\downarrow$ & \LR $\uparrow$ & \ESELD $\downarrow$ \\

\midrule
\multicolumn{6}{l}{\textit{Real-World Only (Imbalance Factor: 552$\times$)}} \\
\midrule

\textbf{MSE}
& 0.616 & 22.0\% & $60.3^\circ$ & 31.2\% & 0.605 \\ 

\textbf{D1}
& \textbf{0.612} & \textbf{31.0}\% & $\textbf{20.7}^\circ$ & \textbf{49.5}\% & \textbf{0.481} \\

\textbf{D2}
& 0.615 & 30.5\% & $21.9^\circ$ & 48.4\% & 0.487 \\

\midrule
\multicolumn{6}{l}{\textit{Real + Synthetic (Imbalance Factor: 17$\times$)}} \\
\midrule

\textbf{MSE}
& \textbf{0.557} & 36.0\% & $19.5^\circ$ & 52.8\% & 0.444 \\ 

\textbf{D1}
& 0.577 & 37.6\% & $19.3^\circ$ & \textbf{62.4}\% & 0.421 \\

\textbf{D2}
& 0.563 & \textbf{38.5}\% & $\textbf{19.2}^\circ$ & 62.3\% & \textbf{0.415} \\

\bottomrule
\end{tabularx}
\end{table}

To further investigate the robustness of the MAGENTA framework, we evaluate its performance when the real-world STARSS23 training set is supplemented with synthetic data (DCASE-Syn), a standard practice in state-of-the-art SELD systems~\cite{shul2024cst, wang2023four}. While this successfully alleviates the absolute scarcity of tail classes, the combined distribution remains imbalanced, and the validation split consists strictly of highly imbalanced real-world recordings. As shown in Table~\ref{tab:sota_dataset}, augmenting the training data naturally improves the standard MSE baseline, lowering \ESELD\ to 0.444. However, MAGENTA consistently outperforms this baseline. The D2 configuration achieves the best overall performance with an \ESELD\ of 0.415, confirming that dynamic difficulty tracking and intrinsic annealing remain highly effective even when absolute data scarcity is mitigated. 

Crucially, the introduction of synthetic data reveals a shift in the optimal angular loss formulation. As demonstrated in Table~\ref{tab:Design_Choices}, the MIA loss (D1) outperforms the perpendicular error loss (D2) when trained exclusively on real-world datasets by forcing spatial learning despite substantial ``detection timidity". However, in the augmented dataset, D2 consistently outperforms D1. This reversal highlights the domain shift inherent in synthetic acoustic data~\cite{yeow2025_enhancing}. While synthetic augmentation provides sufficient samples to overcome the initial timidity, its idealized spatial cues lack the complex coherence and reverberation characteristics of real-world recordings. Because the MIA loss enforces strict angular gradients regardless of detection confidence, it inadvertently causes the network to overfit to these synthetic spatial representations. 

Conversely, the perpendicular error loss inherently acts as a confidence-weighted spatial regularizer. By coupling the angular penalty to the predicted magnitude, $\mathcal{L}_\perp$ permits ``softer" spatial gradients when the model is uncertain. This prevents the network from rigidly overfitting to synthetic spatial cues, promoting superior generalization to the real-world validation split, thereby achieving the highest \Fone\ of 38.5\%. Ultimately, this demonstrates that MAGENTA is a highly modular framework. The MIA loss promotes stronger angular learning for scarce, real-world data, while the perpendicular error loss acts as an essential regularizer for hybrid training conditions.

\subsection{Generalization Across State-of-the-Art Model Architectures}
\label{subsection:generalization_across_sota}

\begin{table*}[th]
  \centering

    \caption{Performance comparison of state-of-the-art SELD model architectures trained with the conventional MSE objective versus the proposed MAGENTA loss framework.}
    \label{tab:sota_Comparison}

    \begin{tabular}{l cc c | cccc c}
    \toprule

    \textbf{Architecture} & \textbf{Params (M)} & \textbf{MACs (G)} &
    \textbf{Loss Objective} &

    \ER $\downarrow$ & \Fone $\uparrow$ & \LE $\downarrow$ & \LR $\uparrow$ & \ESELD $\downarrow$ \\

    \midrule

    \multirow{2}{*}{SELDNet~\cite{adavanne2018sound}} &
    \multirow{2}{*}{0.73} &
    \multirow{2}{*}{0.13} &
    MSE &
    0.557 & 36.0\% & $19.5^\circ$ & 52.8\% & 0.444 \\

    & & & MAGENTA (D2) &
    0.563 & 38.5\% & $19.2^\circ$ & 62.3\% & \textbf{0.415} $(6.5\% \downarrow)$\\

    \midrule
    \multirow{2}{*}{ResNet-biGRU~\cite{niu2023experimental}} &
    \multirow{2}{*}{4.53} &
    \multirow{2}{*}{0.85} & 
    MSE &
    0.541 & 40.1\% & $20.2^\circ$ & 57.5\% & 0.419 \\

    & & & MAGENTA (D2) &
    0.542 & 41.7\% & $19.5^\circ$ & 63.1\% & \textbf{0.401} $(4.3\% \downarrow)$ \\

    \midrule
    
    \multirow{2}{*}{ResNet-Conformer~\cite{niu2023experimental}} &
    \multirow{2}{*}{5.27} &
    \multirow{2}{*}{0.91} & 
    MSE &
    0.527 & 42.5\% & $18.7^\circ$ & 63.3\% & 0.393 \\

    & & & MAGENTA (D2) &
    0.530 & 45.8\% & $17.6^\circ$ & 66.4\% & \textbf{0.376} $(4.3\% \downarrow)$ \\

    \midrule
    
    \multirow{2}{*}{CST-Former~\cite{shul2024cst}} &
    \multirow{2}{*}{0.38} &
    \multirow{2}{*}{1.01} & 
    MSE &
    0.554 & 40.3\% & $19.1^\circ$ & 59.5\% & 0.416 \\

    & & & MAGENTA (D2) &
    0.556 & 41.1\% & $18.5^\circ$ & 64.2\% & \textbf{0.401} $(3.6\% \downarrow)$ \\

    \bottomrule
    
    \end{tabular}
\end{table*}

To demonstrate that the performance gains achieved by MAGENTA are not restricted by a specific neural network backbone, we evaluate its effectiveness across a diverse set of state-of-the-art model architectures. For comparative analysis, the STARSS23 training set is supplemented with the synthetic DCASE-Syn to prevent high-capacity architectures, such as the ResNet-Conformer~\cite{wang2023four, niu2023experimental}, from overfitting to the low-resource real-world data~\cite{wang2023loss}.

Table~\ref{tab:sota_Comparison} details the performance of these models utilizing the conventional MSE objective versus our proposed MAGENTA (D2) configuration. As shown, substituting MSE with MAGENTA yields consistent, significant improvements across all tested architectures. Specifically, the ResNet-Conformer, which serves as the strongest baseline with an MSE-trained \ESELD\ of 0.393, achieves a further 4.3\% relative error reduction using MAGENTA. 

This consistent performance improvement highlights that architectural complexity alone cannot overcome the inherent bias of standard continuous regression objectives under class imbalance. By directly rectifying the optimization landscape, MAGENTA functions as a highly modular, architecture-agnostic framework that provides robust performance gains entirely orthogonal to underlying network design. 

\subsection{Analysis of Computational Overhead}
\label{subsection:comp_overhead}

\begin{table}[t]
    \centering
    \caption{Comparison of computational overhead between the standard MSE baseline and the MAGENTA (D1) framework, benchmarked on a Tesla P40 GPU utilizing a batch size of 64.}
    \label{tab:hardware_compcost}
    
    \begin{tabularx}{\linewidth}{l C C C}
    \toprule
    \textbf{Loss Function} & Avg. Epoch Time (s) & Avg. Loss Latency (ms) & Avg. Inference Time (ms) \\

    \midrule
    MSE Baseline & 19.91 & 8.40 & 3.49 \\

    \midrule
    MAGENTA (D1) & 20.39 & 11.63 & 3.49 \\

    \bottomrule
    \end{tabularx}
\end{table}

To evaluate the practical feasibility of the proposed framework, we benchmark the computational overhead of MAGENTA (D1) against the standard MSE baseline. Table~\ref{tab:hardware_compcost} details the benchmarking results, with all experiments conducted using a Tesla P40 GPU. The geometric decomposition and online difficulty tracking within MAGENTA increase the isolated loss calculation latency from 8.40\,ms to 11.63\,ms per batch. Despite this, however, its impact on the total training cycle is marginal. Specifically, the total training time per epoch only increases by around 2.4\% (19.91\,s to 20.39\,s). 

Notably, the inference latency per pass remains consistent at 3.49\,ms for both configurations. Because the geometric decompositions, rarity priors, and difficulty-driven annealing are strictly training-time operations, MAGENTA incurs zero additional computational cost during inference, making it highly suitable for training models for real-time deployment~\cite{yeow2024real}.

\subsection{Limitations and Future Work}
\label{subsection:limitations_future_work}

While MAGENTA effectively mitigates detection timidity in long-tailed SELD, certain limitations present avenues for future research. 

First, the framework is formulated utilizing single-track ACCDOA~\cite{shimada2021accdoa} to isolate and evaluate the geometric loss decomposition without introducing the confounding assignment complexities of Permutation Invariant Training~\cite{shimada2022multi}. Extending to the multi-ACCDOA format~\cite{shimada2022multi} is essential for resolving same-class overlaps, which will require new mechanisms to track difficulty across permutation-invariant tracks. Furthermore, integrating MAGENTA as a fine-tuning objective for heavily pre-trained architectures (such as PSELDNet~\cite{hu2025pseldnets}) presents a distinct challenge. Because MAGENTA's intrinsic annealing dynamically tracks difficulty based on the model's initial learning trajectory, applying it to networks with pre-existing, robust feature representations would require a substantial redesign of the EMA momentum and scaling mechanics to avoid heavily skewed penalty assignments. Developing such a warm-start initialization protocol represents a concrete direction for future work.

Second, while our current intrinsic difficulty-driven curriculum targets class-wise imbalance, real-world acoustic data can exhibit spatial distribution imbalance~\cite{politis2022starss22, shimada2023starss23}, where sound sources predominantly originate from specific angular regions (e.g., front-facing human speech). Modulating the angular loss based on regional spatial density represents a promising step to prevent spatial overfitting. Furthermore, our unit-sphere decomposition inherently couples vector magnitude to semantic activity confidence. To generalize the MAGENTA framework for full 3-D localization paradigms that include distance estimation~\cite{krause2024sound, yeow2025improving} or motion cues~\cite{yasuda20246dof}, the framework must evolve to mathematically disentangle the physical, distance-based magnitude from semantic activity. Additionally, integrating cross-modal visual cues~\cite{berghi2024fusion} will require redesigning the difficulty tracking to handle altered learning complexities.

Finally, our hybrid training experiments highlighted the persistent challenge of domain shift~\cite{yeow2025_enhancing}. Forcing rigid spatial gradients via $\mathcal{L}_{\mathrm{mia}}$ can overfit to the idealized representations of synthetic data. Future work will explore generalized angular formulations and robust sampling strategies~\cite{kim2025robust} to better bridge the synthetic-to-real domain gap.

\section{Conclusion}

In this paper, we presented MAGENTA, a dynamic, geometrically meaningful loss framework designed to tackle the extreme long-tailed class distributions inherent in real-world SELD. We identified that standard continuous regression objectives critically fail under severe class imbalance, inducing a phenomenon termed ``detection timidity" due to the zero-norm gradient dominance of overwhelmingly inactive frames for rare classes. By geometrically decoupling the ACCDOA error vector into orthogonal radial and angular components, MAGENTA isolates active detection from spatial localization, enabling an online, difficulty-driven learning curriculum that balances tail-class recall with head-class precision. 

Beyond the empirical gains demonstrated on the STARSS23 dataset, including a 20.5\% reduction in the SELD error, this work highlights a broader theoretical implication for machine learning. We demonstrate that treating coupled spatiotemporal representations strictly as generic Euclidean points is suboptimal. Explicit geometric awareness and physical error disentanglement are crucial requirements for robust loss formulations in heavily imbalanced, continuous spaces. 

Moving forward, we aim to extend the geometric decoupling of MAGENTA to the multi-ACCDOA format in order to address overlapping events of the same class in densely polyphonic scenes. Furthermore, we intend to expand our dynamic difficulty tracking to account for spatial, rather than strictly class-wise, imbalances. Ultimately, by establishing a robust and geometrically meaningful framework, this work provides a comprehensive foundation for advancing environmental acoustic intelligence in real-world applications.

\bibliographystyle{IEEEtran}
\bibliography{refs}

@article{adavanne2018sound,
  author={Adavanne, Sharath and Politis, Archontis and Nikunen, Joonas and Virtanen, Tuomas},
  journal={IEEE Journal of Selected Topics in Signal Processing}, 
  title={Sound Event Localization and Detection of Overlapping Sources Using Convolutional Recurrent Neural Networks}, 
  year={2019},
  volume={13},
  number={1},
  pages={34-48},
  doi={10.1109/JSTSP.2018.2885636}}

@article{tang2020long,
  title={Long-tailed classification by keeping the good and removing the bad momentum causal effect},
  author={Tang, Kaihua and Huang, Jianqiang and Zhang, Hanwang},
  journal={Advances in Neural Information Processing Systems},
  volume={33},
  pages={1513--1524},
  year={2020}}

@inproceedings{pu2025leveraging,
  title={Leveraging group classification with descending soft labeling for deep imbalanced regression},
  author={Pu, Ruizhi and Xu, Gezheng and Fang, Ruiyi and Bao, Bing-Kun and Ling, Charles and Wang, Boyu},
  booktitle={Proceedings of the AAAI Conference on Artificial Intelligence},
  volume={39},
  number={19},
  pages={19978--19985},
  year={2025}}

@inproceedings{niu2023experimental,
  author={Niu, Shutong and Du, Jun and Wang, Qing and Chai, Li and Wu, Huaxin and Nian, Zhaoxu and Sun, Lei and Fang, Yi and Pan, Jia and Lee, Chin-Hui},
  booktitle={IEEE International Conference on Acoustics, Speech and Signal Processing (ICASSP)}, 
  title={An Experimental Study on Sound Event Localization and Detection Under Realistic Testing Conditions}, 
  year={2023},
  volume={},
  number={},
  pages={1-5},
  doi={10.1109/ICASSP49357.2023.10094681}}

@inproceedings{wang2021seesaw,
  author={Wang, Jiaqi and Zhang, Wenwei and Zang, Yuhang and Cao, Yuhang and Pang, Jiangmiao and Gong, Tao and Chen, Kai and Liu, Ziwei and Loy, Chen Change and Lin, Dahua},
  booktitle={IEEE/CVF Conference on Computer Vision and Pattern Recognition (CVPR)}, 
  title={Seesaw Loss for Long-Tailed Instance Segmentation}, 
  year={2021},
  volume={},
  number={},
  pages={9690-9699},
  doi={10.1109/CVPR46437.2021.00957}}

@inproceedings{wu2020distribution,
  title={Distribution-Balanced Loss for Multi-Label Classification in Long-Tailed Datasets},
  author={Wu, Tong and Huang, Qingqiu and Liu, Ziwei and Wang, Yu and Lin, Dahua},
  booktitle={European Conference on Computer Vision (ECCV)},
  year={2020}}

@article{yang2022survey,
  title={A survey on long-tailed visual recognition},
  author={Yang, Lu and Jiang, He and Song, Qing and Guo, Jun},
  journal={International Journal of Computer Vision},
  volume={130},
  number={7},
  pages={1837--1872},
  year={2022},
  publisher={Springer}}

@article{yeow2025environmental,
  title={Environmental acoustic intelligence through sound event localization and detection: a review},
  author={Yeow, Jun-Wei and Tan, Ee-Leng and Peksi, Santi and Gan, Woon-Seng},
  journal={npj Acoustics},
  volume={1},
  number={1},
  pages={31},
  year={2025},
  publisher={Nature Publishing Group UK London}}

@inproceedings{cui2019class,
  title={Class-balanced loss based on effective number of samples},
  author={Cui, Yin and Jia, Menglin and Lin, Tsung-Yi and Song, Yang and Belongie, Serge},
  booktitle={Proceedings of the IEEE/CVF Conference on Computer Vision and Pattern Recognition},
  pages={9268--9277},
  year={2019}}

@inproceedings{ren2022balanced,
  title={Balanced MSE for Imbalanced Visual Regression},
  author={Ren, Jiawei and Zhang, Mingyuan and Yu, Cunjun and Liu, Ziwei},
  booktitle={Proceedings of the IEEE/CVF Conference on Computer Vision and Pattern Recognition},
  year={2022}}

@inproceedings{suzic2024exterior,
  author={Suzić, Siniša and Martín-Morató, Irene and Simić, Nikola and Raghavaraju, Charitha and Heittola, Toni and Stanojev, Vuk and Bajovic, Dragana},
  booktitle={32nd European Signal Processing Conference (EUSIPCO)}, 
  title={UNS Exterior Spatial Sound Events Dataset for Urban Monitoring}, 
  year={2024},
  volume={},
  number={},
  pages={176-180},
  doi={10.23919/EUSIPCO63174.2024.10715448}}

@inproceedings{yasuda20246dof,
  author={Yasuda, Masahiro and Saito, Shoichiro and Nakayama, Akira and Harada, Noboru},
  booktitle={IEEE International Conference on Acoustics, Speech and Signal Processing (ICASSP)}, 
  title={6DoF SELD: Sound Event Localization and Detection Using Microphones and Motion Tracking Sensors on Self-Motioning Human}, 
  year={2024},
  volume={},
  number={},
  pages={1411-1415},
  doi={10.1109/ICASSP48485.2024.10446749}}

@techreport{kim2025robust,
    Author = "Kim, Jin Sob and Park, Hyun Joon and Shin, Wooseok and Han, Sung Won",
    title = "A ROBUST FRAMEWORK FOR SOUND EVENT LOCALIZATION AND DETECTION ON REAL RECORDINGS",
    institution = "DCASE2022 Challenge",
    year = "2022",
    month = "June"}

@inproceedings{ridnik2021asymmetric,
  author={Ridnik, Tal and Ben-Baruch, Emanuel and Zamir, Nadav and Noy, Asaf and Friedman, Itamar and Protter, Matan and Zelnik-Manor, Lihi},
  booktitle={IEEE/CVF International Conference on Computer Vision (ICCV)}, 
  title={Asymmetric Loss For Multi-Label Classification}, 
  year={2021},
  volume={},
  number={},
  pages={82-91},
  doi={10.1109/ICCV48922.2021.00015}}

@article{politis2020overview,
  title={Overview and evaluation of sound event localization and detection in DCASE 2019},
  author={Politis, Archontis and Mesaros, Annamaria and Adavanne, Sharath and Heittola, Toni and Virtanen, Tuomas},
  journal={IEEE/ACM Transactions on Audio, Speech, and Language Processing},
  volume={29},
  pages={684--698},
  year={2020},
  publisher={IEEE}}

@article{shimada2023starss23,
  title={STARSS23: An audio-visual dataset of spatial recordings of real scenes with spatiotemporal annotations of sound events},
  author={Shimada, Kazuki and Politis, Archontis and Sudarsanam, Parthasaarathy and Krause, Daniel A and Uchida, Kengo and Adavanne, Sharath and Hakala, Aapo and Koyama, Yuichiro and Takahashi, Naoya and Takahashi, Shusuke and others},
  journal={Advances in Neural Information Processing Systems},
  volume={36},
  pages={72931--72957},
  year={2023}}

@article{Tan_EqualizationLosses_2023,
  author={Tan, Jingru and Li, Bo and Lu, Xin and Yao, Yongqiang and Yu, Fengwei and He, Tong and Ouyang, Wanli},
  journal={IEEE Transactions on Pattern Analysis and Machine Intelligence}, 
  title={The Equalization Losses: Gradient-Driven Training for Long-tailed Object Recognition}, 
  year={2023},
  volume={45},
  number={11},
  pages={13876-13892},
  doi={10.1109/TPAMI.2023.3298433}}

@inproceedings{roman2024spatial,
  author={Roman, Iran R. and Ick, Christopher and Ding, Sivan and Roman, Adrian S. and McFee, Brian and Bello, Juan P.},
  booktitle={IEEE International Conference on Acoustics, Speech and Signal Processing (ICASSP)}, 
  title={Spatial Scaper: A Library to Simulate and Augment Soundscapes for Sound Event Localization and Detection in Realistic Rooms}, 
  year={2024},
  volume={},
  number={},
  pages={1221-1225},
  doi={10.1109/ICASSP48485.2024.10446118}}

@inproceedings{shimada2025stereo,
    author = "Shimada, Kazuki and Politis, Archontis and Roman, Iran and Sudarsanam, Parthasaarathy and Diaz-Guerra, David and Pandey, Ruchi and Uchida, Kengo and Koyama, Yuichiro and Takahashi, Naoya and Shibuya, Takashi and Takahashi, Shusuke and Virtanen, Tuomas and Mitsufuji, Yuki",
    title = "Stereo Sound Event Localization and Detection with Onscreen/Offscreen Classification",
    booktitle = "Proceedings of the 10th Workshop on Detection and Classification of Acoustic Scenes and Events (DCASE 2025)",
    address = "Barcelona, Spain",
    month = "October",
    year = "2025",
    pages = "140--144",
    isbn = "978-84-09-77652-8",
    doi = "10.5281/zenodo.17251589"}

@article{yeow2024real,
  title={Real-time sound event localization and detection: Deployment challenges on edge devices},
  author={Yeow, Jun Wei and Tan, Ee-Leng and Bai, Jisheng and Peksi, Santi and Gan, Woon-Seng},
  journal={arXiv preprint arXiv:2409.11700},
  year={2024}}

@techreport{yeow2025improving,
    Author = "Yeow, Jun-Wei and Tan, Ee-Leng and Peksi, Santi and Gan, Woon-Seng",
    title = "Improving stereo 3d sound event localization and detection: perceptual features, stereo-specific data augmentation, and distance normalization",
    institution = "DCASE2025 Challenge",
    year = "2025",
    month = "June"}

@article{yeow2025_enhancing,
  author={Yeow, Jun-Wei and Tan, Ee-Leng and Bai, Jisheng and Peksi, Santi and Gan, Woon-Seng},
  journal={IEEE Sensors Journal}, 
  title={Enhancing 3-D Sound Event Localization and Detection With Distance Estimation Using Reverberation and Spatial Coherence Features}, 
  year={2025},
  volume={25},
  number={15},
  pages={29221-29237},
  doi={10.1109/JSEN.2025.3583033}}

@article{zhang2023deep,
  author={Zhang, Yifan and Kang, Bingyi and Hooi, Bryan and Yan, Shuicheng and Feng, Jiashi},
  journal={IEEE Transactions on Pattern Analysis and Machine Intelligence}, 
  title={Deep Long-Tailed Learning: A Survey}, 
  year={2023},
  volume={45},
  number={9},
  pages={10795-10816},
  doi={10.1109/TPAMI.2023.3268118}}

@article{lin2017focal,
  author={Lin, Tsung-Yi and Goyal, Priya and Girshick, Ross and He, Kaiming and Dollár, Piotr},
  journal={IEEE Transactions on Pattern Analysis and Machine Intelligence}, 
  title={Focal Loss for Dense Object Detection}, 
  year={2020},
  volume={42},
  number={2},
  pages={318-327},
  doi={10.1109/TPAMI.2018.2858826}}

@techreport{yeow2024squeeze,
    Author = "Yeow, Jun Wei and Tan, Ee-Leng and Bai, Jisheng and Peksi, Santi and Gan, Woon-Seng",
    title = "SQUEEZE-AND-EXCITE RESNET-CONFORMERS FOR SOUND EVENT LOCALIZATION, DETECTION, AND DISTANCE ESTIMATION FOR DCASE2024 CHALLENGE",
    institution = "DCASE2024 Challenge",
    year = "2024",
    month = "June"}

@inproceedings{imoto2021impact,
  author={Imoto, Keisuke and Mishima, Sakiko and Arai, Yumi and Kondo, Reishi},
  booktitle={IEEE International Conference on Acoustics, Speech and Signal Processing (ICASSP)}, 
  title={Impact of Sound Duration and Inactive Frames on Sound Event Detection Performance}, 
  year={2021},
  volume={},
  number={},
  pages={860-864},
  doi={10.1109/ICASSP39728.2021.9414949}}

@inproceedings{shul2024cst,
  author={Shul, Yusun and Choi, Jung-Woo},
  booktitle={IEEE International Conference on Acoustics, Speech and Signal Processing (ICASSP)}, 
  title={CST-Former: Transformer with Channel-Spectro-Temporal Attention for Sound Event Localization and Detection}, 
  year={2024},
  volume={},
  number={},
  pages={8686-8690},
  doi={10.1109/ICASSP48485.2024.10447181}}

@article{imoto2022impact,
  title={Impact of data imbalance caused by inactive frames and difference in sound duration on sound event detection performance},
  author={Imoto, Keisuke and Mishima, Sakiko and Arai, Yumi and Kondo, Reishi},
  journal={Applied Acoustics},
  volume={196},
  pages={108882},
  year={2022},
  publisher={Elsevier}}

@inproceedings{wang2023loss,
  author={Wang, Qing and Du, Jun and Nian, Zhaoxu and Niu, Shutong and Chai, Li and Wu, Huaxin and Pan, Jia and Lee, Chin-Hui},
  booktitle={IEEE International Conference on Acoustics, Speech and Signal Processing (ICASSP)}, 
  title={Loss Function Design for DNN-Based Sound Event Localization and Detection on Low-Resource Realistic Data}, 
  year={2023},
  volume={},
  number={},
  pages={1-5},
  doi={10.1109/ICASSP49357.2023.10095144}}

@inproceedings{shimada2021accdoa,
  author={Shimada, Kazuki and Koyama, Yuichiro and Takahashi, Naoya and Takahashi, Shusuke and Mitsufuji, Yuki},
  booktitle={IEEE International Conference on Acoustics, Speech and Signal Processing (ICASSP)}, 
  title={Accdoa: Activity-Coupled Cartesian Direction of Arrival Representation for Sound Event Localization And Detection}, 
  year={2021},
  volume={},
  number={},
  pages={915-919},
  doi={10.1109/ICASSP39728.2021.9413609}}

@inproceedings{shimada2022multi,
  author={Shimada, Kazuki and Koyama, Yuichiro and Takahashi, Shusuke and Takahashi, Naoya and Tsunoo, Emiru and Mitsufuji, Yuki},
  booktitle={IEEE International Conference on Acoustics, Speech and Signal Processing (ICASSP)}, 
  title={Multi-ACCDOA: Localizing And Detecting Overlapping Sounds From The Same Class With Auxiliary Duplicating Permutation Invariant Training}, 
  year={2022},
  volume={},
  number={},
  pages={316-320},
  doi={10.1109/ICASSP43922.2022.9746384}}

@inproceedings{krause2024sound,
  author={Krause, Daniel Aleksander and Politis, Archontis and Mesaros, Annamaria},
  booktitle={32nd European Signal Processing Conference (EUSIPCO)}, 
  title={Sound Event Detection and Localization with Distance Estimation}, 
  year={2024},
  volume={},
  number={},
  pages={286-290},
  doi={10.23919/EUSIPCO63174.2024.10715220}}

@inproceedings{zhang2024unified,
  author={Zhang, Yuliang and Togneri, Roberto and Huang, David},
  booktitle={IEEE International Conference on Acoustics, Speech and Signal Processing (ICASSP)}, 
  title={A Unified Loss Function to Tackle Inter-Class and Intra-Class Data Imbalance in Sound Event Detection}, 
  year={2024},
  volume={},
  number={},
  pages={996-1000},
  doi={10.1109/ICASSP48485.2024.10447675}}

@inproceedings{cao2019polyphonic,
    author = "Cao, Yin and Kong, Qiuqiang and Iqbal, Turab and An, Fengyan and Wang, Wenwu and Plumbley, Mark",
    title = "Polyphonic Sound Event Detection and Localization using a Two-Stage Strategy",
    booktitle = "Proceedings of the Detection and Classification of Acoustic Scenes and Events 2019 Workshop (DCASE2019)",
    address = "New York University, NY, USA",
    month = "October",
    year = "2019",
    pages = "30--34"}

@inproceedings{cao2019LDAM,
  title={Learning Imbalanced Datasets with Label-Distribution-Aware Margin Loss},
  author={Cao, Kaidi and Wei, Colin and Gaidon, Adrien and Arechiga, Nikos and Ma, Tengyu},
  booktitle={Advances in Neural Information Processing Systems},
  year={2019}}

@article{li2022generalizedfocal,
  author={Li, Xiang and Lv, Chengqi and Wang, Wenhai and Li, Gang and Yang, Lingfeng and Yang, Jian},
  journal={IEEE Transactions on Pattern Analysis and Machine Intelligence}, 
  title={Generalized Focal Loss: Towards Efficient Representation Learning for Dense Object Detection}, 
  year={2023},
  volume={45},
  number={3},
  pages={3139-3153},
  doi={10.1109/TPAMI.2022.3180392}}

@article{wang2023four,
  author={Wang, Qing and Du, Jun and Wu, Hua-Xin and Pan, Jia and Ma, Feng and Lee, Chin-Hui},
  journal={IEEE/ACM Transactions on Audio, Speech, and Language Processing}, 
  title={A Four-Stage Data Augmentation Approach to ResNet-Conformer Based Acoustic Modeling for Sound Event Localization and Detection}, 
  year={2023},
  volume={31},
  number={},
  pages={1251-1264},
  doi={10.1109/TASLP.2023.3256088}}

@article{chawla2002smote,
  title={SMOTE: synthetic minority over-sampling technique},
  author={Chawla, Nitesh V and Bowyer, Kevin W and Hall, Lawrence O and Kegelmeyer, W Philip},
  journal={Journal of Artificial Intelligence Research},
  volume={16},
  pages={321--357},
  year={2002}}

@book{virtanen2018computational,
  title={Computational analysis of sound scenes and events},
  author={Virtanen, Tuomas and Plumbley, Mark D and Ellis, Dan},
  volume={9},
  year={2018},
  publisher={Springer}}

@article{zhang2025systematic,
  author={Zhang, Chongsheng and Almpanidis, George and Fan, Gaojuan and Deng, Binquan and Zhang, Yanbo and Liu, Ji and Kamel, Aouaidjia and Soda, Paolo and Gama, João},
  journal={IEEE Transactions on Neural Networks and Learning Systems}, 
  title={A Systematic Review on Long-Tailed Learning}, 
  year={2025},
  volume={36},
  number={8},
  pages={13670-13690},
  doi={10.1109/TNNLS.2025.3539314}}

@inproceedings{politis2022starss22,
    author = "Politis, Archontis and Shimada, Kazuki and Sudarsanam, Parthasaarathy and Adavanne, Sharath and Krause, Daniel and Koyama, Yuichiro and Takahashi, Naoya and Takahashi, Shusuke and Mitsufuji, Yuki and Virtanen, Tuomas",
    title = "{STARSS22}: {A} dataset of spatial recordings of real scenes with spatiotemporal annotations of sound events",
    booktitle = "Proceedings of the 8th Detection and Classification of Acoustic Scenes and Events 2022 Workshop (DCASE2022)",
    address = "Nancy, France",
    month = "November",
    year = "2022",
    pages = "125--129"}

@inproceedings{zhang2022label,
  author={Zhang, Shaoyu and Chen, Chen and Zhang, Xiujuan and Peng, Silong},
  booktitle={IEEE International Conference on Acoustics, Speech and Signal Processing (ICASSP)}, 
  title={Label-Occurrence-Balanced Mixup for Long-Tailed Recognition}, 
  year={2022},
  volume={},
  number={},
  pages={3224-3228},
  doi={10.1109/ICASSP43922.2022.9746299}}

@inproceedings{yeow2025towards,
  author={Yeow, Jun-Wei and Tan, Ee-Leng and Peksi, Santi and Gan, Woon-Seng and Huang, Qirui},
  booktitle={Asia Pacific Signal and Information Processing Association Annual Summit and Conference (APSIPA ASC)}, 
  title={Towards Robust Stereo 3-D SELD: A Study of Perceptual Features and Data Augmentation}, 
  year={2025},
  volume={},
  number={},
  pages={95-100},
  doi={10.1109/APSIPAASC65261.2025.11249371}}

@inproceedings{scheibler2018pyroomacoustics,
  author={Scheibler, Robin and Bezzam, Eric and Dokmanić, Ivan},
  booktitle={2018 IEEE International Conference on Acoustics, Speech and Signal Processing (ICASSP)}, 
  title={Pyroomacoustics: A Python Package for Audio Room Simulation and Array Processing Algorithms}, 
  year={2018},
  volume={},
  number={},
  pages={351-355},
  doi={10.1109/ICASSP.2018.8461310}}

@article{zhang2025sound,
  author={Zhang, Dongzhe and Chen, Jianfeng and Bai, Jisheng and Wang, Mou and Shi, Dongyuan and Niu, Qixiang and Bernardini, Alberto},
  journal={IEEE Sensors Journal}, 
  title={Sound Event Localization and Classification Using Wireless Acoustic Sensor Networks in Outdoor Environments}, 
  year={2025},
  volume={25},
  number={22},
  pages={42141-42153},
  doi={10.1109/JSEN.2025.3614680}}

@inproceedings{tian2022striking,
  title={Striking the right balance: Recall loss for semantic segmentation},
  author={Tian, Junjiao and Mithun, Niluthpol Chowdhury and Seymour, Zachary and Chiu, Han-Pang and Kira, Zsolt},
  booktitle={2022 International Conference on Robotics and Automation (ICRA)},
  pages={5063--5069},
  year={2022},
  organization={IEEE}}

@inproceedings{mazzon2019first,
    author = "Mazzon, Luca and Koizumi, Yuma and Yasuda, Masahiro and Harada, Noboru",
    title = "First Order Ambisonics Domain Spatial Augmentation for DNN-based Direction of Arrival Estimation",
    booktitle = "Proceedings of the Detection and Classification of Acoustic Scenes and Events 2019 Workshop (DCASE2019)",
    address = "New York University, NY, USA",
    month = "October",
    year = "2019",
    pages = "154--158"}

@inproceedings{perotin2018crnn,
  author={Perotin, Lauréline and Serizel, Romain and Vincent, Emmanuel and Guérin, Alexandre},
  booktitle={16th International Workshop on Acoustic Signal Enhancement (IWAENC)}, 
  title={CRNN-based Joint Azimuth and Elevation Localization with the Ambisonics Intensity Vector}, 
  year={2018},
  volume={},
  number={},
  pages={241-245},
  doi={10.1109/IWAENC.2018.8521403}}

@inproceedings{mesaros2025decade,
  author={Mesaros, Annamaria and Serizel, Romain and Heittola, Toni and Virtanen, Tuomas and Plumbley, Mark D.},
  booktitle={IEEE International Conference on Acoustics, Speech and Signal Processing (ICASSP)}, 
  title={A decade of DCASE: Achievements, practices, evaluations and future challenges}, 
  year={2025},
  volume={},
  number={},
  pages={1-5},
  doi={10.1109/ICASSP49660.2025.10887673}}

@inproceedings{politis2021dataset_tnsse2021,
    author = "Politis, Archontis and Adavanne, Sharath and Krause, Daniel and Deleforge, Antoine and Srivastava, Prerak and Virtanen, Tuomas",
    title = "A Dataset of Dynamic Reverberant Sound Scenes with Directional Interferers for Sound Event Localization and Detection",
    booktitle = "Proceedings of the 6th Detection and Classification of Acoustic Scenes and Events 2021 Workshop (DCASE2021)",
    address = "Barcelona, Spain",
    month = "November",
    year = "2021",
    pages = "125--129"}

@inproceedings{smith2019super,
  title={Super-convergence: Very fast training of neural networks using large learning rates},
  author={Smith, Leslie N and Topin, Nicholay},
  booktitle={Artificial Intelligence and Machine Learning for Multi-Domain Operations Applications},
  volume={11006},
  pages={369--386},
  year={2019},
  organization={SPIE}}

@inproceedings{adavanne2019_tnsse2019,
    author = "Adavanne, Sharath and Politis, Archontis and Virtanen, Tuomas",
    title = "A Multi-room Reverberant Dataset for Sound Event Localization and Detection",
    booktitle = "Proceedings of the Detection and Classification of Acoustic Scenes and Events 2019 Workshop (DCASE2019)",
    address = "New York University, NY, USA",
    month = "October",
    year = "2019",
    pages = "10--14"}

@inproceedings{politis2020_tnsse2020,
    author = "Politis, Archontis and Adavanne, Sharath and Virtanen, Tuomas",
    title = "A Dataset of Reverberant Spatial Sound Scenes with Moving Sources for Sound Event Localization and Detection",
    booktitle = "Proceedings of the Detection and Classification of Acoustic Scenes and Events 2020 Workshop (DCASE2020)",
    address = "Tokyo, Japan",
    month = "November",
    year = "2020",
    pages = "165--169"}

@inproceedings{hirvonen2015classification,
  title={Classification of spatial audio location and content using convolutional neural networks},
  author={Hirvonen, Toni},
  booktitle={Audio Engineering Society Convention 138},
  year={2015},
  organization={Audio Engineering Society}}

@inproceedings{nagatomo2022wearable,
  author={Nagatomo, Kento and Yasuda, Masahiro and Yatabe, Kohei and Saito, Shoichiro and Oikawa, Yasuhiro},
  booktitle={IEEE International Conference on Acoustics, Speech and Signal Processing (ICASSP)}, 
  title={Wearable Seld Dataset: Dataset For Sound Event Localization And Detection Using Wearable Devices Around Head}, 
  year={2022},
  volume={},
  number={},
  pages={156-160},
  doi={10.1109/ICASSP43922.2022.9746544}}

@inproceedings{phan2020multitask,
    author = "Phan, Huy and Pham, Lam and Koch, Philipp and Duong, Ngoc Q. K. and McLoughlin, Ian and Mertins, Alfred",
    title = "On Multitask Loss Function for Audio Event Detection and Localization",
    booktitle = "Proceedings of the Detection and Classification of Acoustic Scenes and Events 2020 Workshop (DCASE2020)",
    address = "Tokyo, Japan",
    month = "November",
    year = "2020",
    pages = "160--164"}

@article{hu2025pseldnets,
  author={Hu, Jinbo and Cao, Yin and Wu, Ming and Kang, Fang and Yang, Feiran and Wang, Wenwu and Plumbley, Mark D. and Yang, Jun},
  journal={IEEE Transactions on Audio, Speech and Language Processing}, 
  title={PSELDNets: Pre-Trained Neural Networks on a Large-Scale Synthetic Dataset for Sound Event Localization and Detection}, 
  year={2025},
  volume={33},
  number={},
  pages={2845-2860},
  doi={10.1109/TASLPRO.2025.3587446}}

@inproceedings{berghi2024fusion,
  author={Berghi, Davide and Wu, Peipei and Zhao, Jinzheng and Wang, Wenwu and Jackson, Philip J. B.},
  booktitle={IEEE International Conference on Acoustics, Speech and Signal Processing (ICASSP)}, 
  title={Fusion of Audio and Visual Embeddings for Sound Event Localization and Detection}, 
  year={2024},
  volume={},
  number={},
  pages={8816-8820},
  doi={10.1109/ICASSP48485.2024.10448050}}

@article{shi2023re,
  title={How re-sampling helps for long-tail learning?},
  author={Shi, Jiang-Xin and Wei, Tong and Xiang, Yuke and Li, Yu-Feng},
  journal={Advances in Neural Information Processing Systems},
  volume={36},
  pages={75669--75687},
  year={2023}}

@inproceedings{yang2021delving,
  title={Delving into Deep Imbalanced Regression},
  author={Yang, Yuzhe and Zha, Kaiwen and Chen, Ying-Cong and Wang, Hao and Katabi, Dina},
  booktitle={International Conference on Machine Learning (ICML)},
  year={2021}}

@inproceedings{park2021influence,
  title={Influence-balanced loss for imbalanced visual classification},
  author={Park, Seulki and Lim, Jongin and Jeon, Younghan and Choi, Jin Young},
  booktitle={Proceedings of the IEEE/CVF international conference on computer vision},
  pages={735--744},
  year={2021}}

\begin{IEEEbiography}[{\includegraphics[width=1in,height=1.25in,clip,keepaspectratio]{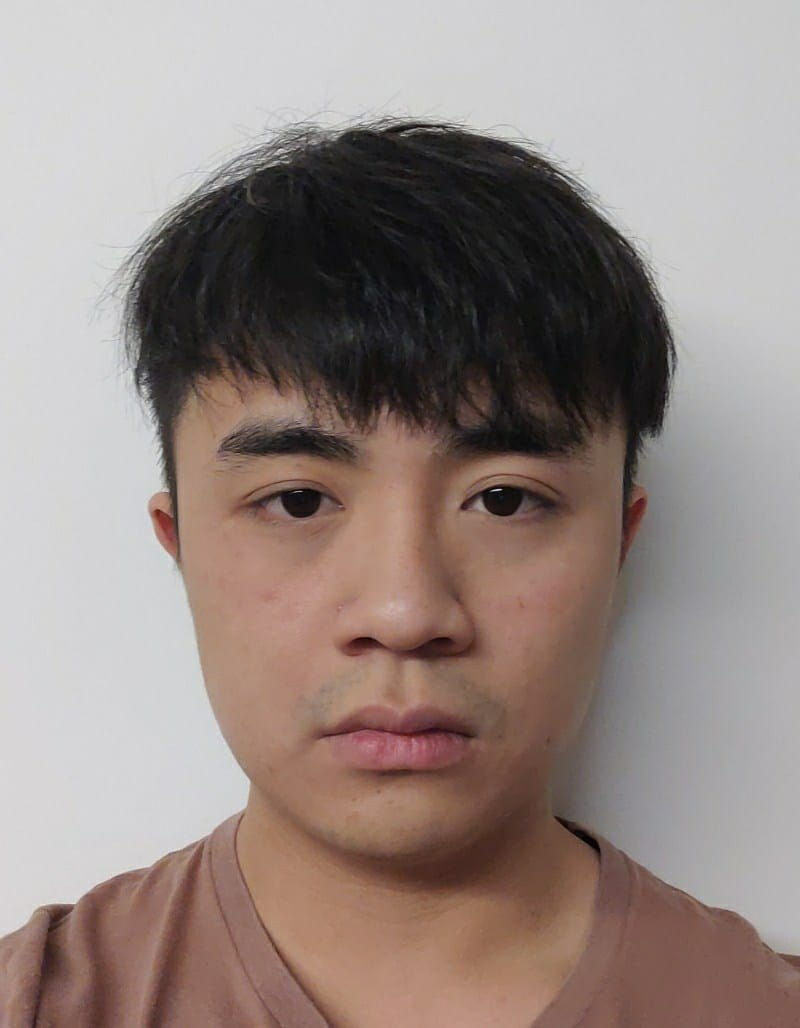}}]{Jun-Wei Yeow} received the B.Eng.\ degree (First-Class Honors) in Electrical and Electronic Engineering from Nanyang Technological University, Singapore, in 2023. He is currently pursuing the Ph.D.\ degree at the same institution. His research interests include sound event localization and detection, machine learning, and real-time signal processing applications.
\end{IEEEbiography}

\begin{IEEEbiography}[{\includegraphics[width=1in,height=1.25in,clip,keepaspectratio]{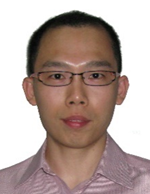}}]{Ee-Leng Tan} received his B.Eng. (1st Class Hons) and Ph.D. degrees, both in Electrical and Electronic Engineering from the Nanyang Technological University (NTU) in 2003 and 2012, respectively. He is currently a Senior Research Fellow at NTU. His research interests include real-time signal processing, perceptual image processing, 3-D and directional audio, and application of machine learning in medical imaging. His work has resulted in three patents.
\end{IEEEbiography}

\begin{IEEEbiography}[{\includegraphics[width=1in,height=1.25in,clip,keepaspectratio]{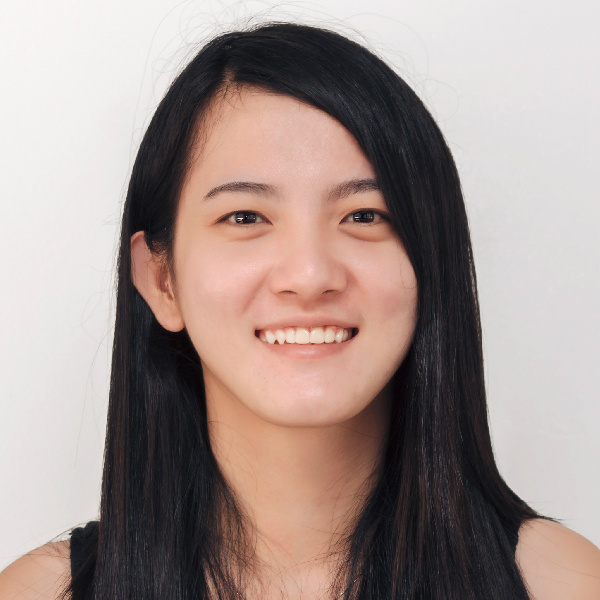}}]{Santi Peksi} received her B.Eng. degree in Engineering in Information Engineering and Media (Second-Class Honors) from Nanyang Technological University (NTU), Singapore, in 2013. She is currently serving as a Senior Research Engineer at NTU. Her research interests encompass real-time signal processing, 3-D audio measurement and rendering, sound event localization and detection, as well as applications of machine learning in acoustic and audio technologies.
\end{IEEEbiography}

\begin{IEEEbiography}[{\includegraphics[width=1in,height=1.25in,clip,keepaspectratio]{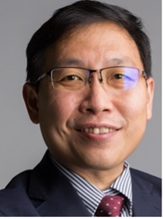}}]{Woon-Seng Gan} is a professor at Nanyang Technological University's School of Electrical and Electronic Engineering and serving as the Director of the Smart Nation TRANS Lab @ NTU in Singapore. Professor Gan holds several prestigious titles, including Fellow of the Audio Engineering Society (AES), Fellow of the Institute of Engineering and Technology (IET), and Senior Member of the IEEE. He has also been selected as the IEEE Signal Processing Society Distinguished Lecturer for 2023-2024. He has published over 400 papers in international refereed journals and conferences.
Professor Gan has contributed significantly to leading journals as an Associate Editor, including the IEEE/ACM Transactions on Audio, Speech, and Language Processing; and the Journal of Audio Engineering Society. He is currently a Senior Area Editor for the IEEE Signal Processing Letters. He has also served as the lead guest editor in various areas, including Active and Passive Noise Control, Parametric Acoustic Array, Spatial Audio, and Intelligent Audio, Speech, and Music Processing. He actively participated in the IEEE Signal Processing Industry DSP committee and the IEEE IoT special interest group and has served on the IEEE SPS Technical Directions Board and Education Board. Professor Gan has been recognized as one of the ``World's Top 2\% Scientists 2021-23" in the field of Acoustics by Stanford University, acknowledging his outstanding achievements. He served as the Technical Program Chair for the first hybrid IEEE International Conference on Acoustics, Speech, and Signal Processing held in Singapore in May 2022. He is currently the distinguished lecturer of the IEEE Signal Processing Society and the President of the Asia Pacific Signal and Information Processing Association (APSIPA) from 2025-2026.

\end{IEEEbiography}

\end{document}